\documentclass[
  twocolumn, aps, prb, showpacs, amsmath,
  amssymb, superscriptaddress, bibnotes,
  ]{revtex4-1} 

\usepackage{hyperref}
\usepackage{graphicx}
\usepackage{mathtools}
\usepackage{hyperref}
\usepackage{color}
\usepackage[usenames,dvipsnames,svgnames,table]{xcolor}

\usepackage[export]{adjustbox}

\newcommand{\mbf}[1]{\mathbf{#1}}

\newcommand{\n}{\hat{n}}

\newcommand{\astcycl}{\mathrlap{\kern0.085em{\circlearrowright}}\ast}

\newcommand{\taucycl}{\mathrlap{\kern0.42em{\bullet}}\circlearrowright}

\newcommand{\ca}{c^{\phantom{\dagger}}}
\newcommand{\cc}{c^\dagger}

\begin{document}

\title{Hund's coupling driven photo-carrier relaxation in the two-band Mott insulator}

\author{Hugo U.~R.~Strand}
\email{hugo.strand@unifr.ch}
\affiliation{Department of Quantum Matter Physics, University of Geneva, 24 Quai Ernest-Ansermet, 1211 Geneva 4, Switzerland}
\affiliation{Department of Physics, University of Fribourg, 1700 Fribourg, Switzerland} 

\author{Denis Gole\v{z}}
\affiliation{Department of Physics, University of Fribourg, 1700 Fribourg, Switzerland} 

\author{Martin Eckstein}
\affiliation{Max Planck Research Department for Structural Dynamics, University of Hamburg-CFEL, 22761 Hamburg, Germany} 

\author{Philipp Werner}
\email{philipp.werner@unifr.ch}
\affiliation{Department of Physics, University of Fribourg, 1700 Fribourg, Switzerland} 

\date{\today} 
\pacs{71.10.Fd, 05.70.Ln}

\begin{abstract}
We study the relaxation dynamics of photo-carriers in the paramagnetic Mott insulating phase of the half-filled two-band Hubbard model.  Using nonequilibrium dynamical mean field theory, we excite charge carriers across the Mott gap by a short hopping modulation, and simulate the evolution of the photo-doped population within the Hubbard bands.
We observe an ultrafast charge-carrier relaxation driven by emission of local spin excitations with an inverse relaxation time proportional to the Hund's coupling.
The photo-doping generates additional side-bands in the spectral function, and for strong Hund's coupling, the photo-doped population also splits into several resonances.
The dynamics of the local many-body states reveals two effects, \emph{thermal blocking} and \emph{kinetic freezing}, which manifest themselves when the Hund's coupling becomes of the order of the temperature or the bandwidth, respectively.
These effects, which are absent in the single-band Hubbard model, should be relevant for the interpretation of experiments on correlated materials with multiple active orbitals.
In particular, the features revealed in the non-equilibrium energy distribution of the photo-carriers are experimentally accessible, and provide information on the role of the Hund's coupling in these materials. 
\end{abstract}

\maketitle
\makeatletter
\let\toc@pre\relax
\let\toc@post\relax
\makeatother


\section{Introduction}


The photo-excitation of charge carriers across a Mott or charge-transfer gap triggers a nonequilibrium phase transition from a correlation-induced insulating to a nonthermal conducting state.
A wide range of experiments have characterized these photo-doped metallic states and explored the relaxation pathways and lifetimes of the photo-carriers, as well as their effect on ordered states.
%
%
Iwai and collaborators \cite{Iwai:2003aa} measured the reflectivity spectrum of a nickel-chain compound, and found that it exhibits a Drude-like low-energy feature immediately after photo-excitation, and that this metallic behavior lasts for several picoseconds.
Similar results have also been obtained for the cuprates La$_2$CuO$_4$ and Nd$_2$CuO$_4$ \cite{Okamoto:2010aa}, where it was also shown that charge-spin and charge-phonon couplings play an important role in the short-time relaxation process.
Using time-resolved photo-emission spectroscopy, the ultrafast relaxation of photo-doped doublons has been investigated in the polaronic Mott insulator 1T-TaS$_2$ \cite{Perfetti:2006aa, Perfetti:2008aa, Ligges:2017aa}, and femtosecond resonant X-ray diffraction has been employed to study the effect of photo-doping on the magnetic structure in CuO \cite{Johnson:2012aa} and TbMnO$_3$ \cite{Johnson:2015aa}. The latter material exhibits a complex interplay between charge, spin, and orbital degrees of freedom.  

Unveiling the nature of the bosonic excitations coupled to the electronic degrees of freedom is one of the fundamental goals of pump-probe spectroscopy.
The two-temperature model introduced by Allen \cite{Allen:1987aa} provides a means to extract the microscopical parameters describing electron-phonon coupling from the relaxation dynamics after photo-excitation \cite{Brorson:1990aa, Gadermaier:2010aa}.
A careful analysis of the time-dependent optical spectroscopy \cite{Heumen:2009aa} and angle-resolved photo-emission spectroscopy (ARPES) measurements \cite{Rameau:2016aa} enabled the determination of the relative weight of the electron-boson coupling for a broad range of frequencies \cite{Dal-Conte:2012aa, Dal-Conte:2015aa}.
The population dynamics measured in ARPES furthermore allows to disentangle the electron-electron interactions from electron-boson interactions \cite{Rameau:2016aa}.
This information is crucial for determining the microscopic origin of symmetry broken states, such as superconductivity \cite{Dal-Conte:2015aa}, density-wave orders \cite{Porer:2014aa}, excitonic insulators, etc., especially when several orders are intertwined \cite{Fradkin:2015aa}.

Significant theoretical effort has also been aimed at understanding the basic aspects of the photo-doping process and the subsequent relaxation \cite{Aoki:2014kx, Kemper:2016aa, Strohmaier:2010aa}.
These studies -- which so far have focused on single-band models --
have addressed the exponential dependence of the photo-carrier's life-time on the Mott gap \cite{Eckstein:2011aa, Lenarifmmode-celse-cfiiifmmode-celse-cfi:2013aa, Strohmaier:2010aa},
the role of impact ionization processes in small-gap insulators \cite{Werner:2014aa}, and the nature of the photo-doped metallic state \cite{Eckstein:2013aa}.
Deep in the Mott phase the number of holon-doublon pairs is an almost conserved quantity on the timescale of the electronic hopping, and the initial relaxation of the photo carriers can only occur within the Hubbard bands.
The corresponding relaxation time strongly depends on the coupling of the photo-doped carriers to bosonic degrees of freedom, such as spins \cite{Werner:2012aa, Goleifmmode-zelse-zfi:2015ab} or phonons \cite{0295-5075-109-3-37002, Dorfner:2015aa, Sayyad:2015aa, Murakami:2015aa, Goleifmmode-zelse-zfi:2012aa}.
In particular the scattering with antiferromagnetically ordered spins provides a very efficient cooling mechanism \cite{Werner:2012aa, Eckstein:2016aa, Goleifmmode-zelse-zfi:2014aa, Dal-Conte:2015aa}. 
Very strong electron-boson couplings, which appear for example in systems with dynamically screened Coulomb interactions \cite{Golez:2017aa}, can on the other hand open new channels for doublon-hole recombination and lead to a nontrivial energy exchange between electronic and bosonic degrees of freedom \cite{0295-5075-109-3-37002, Golez:2017aa, doi:10.1063/1.4935245}.

%
%
\begin{figure}
  \includegraphics[scale=1]
  {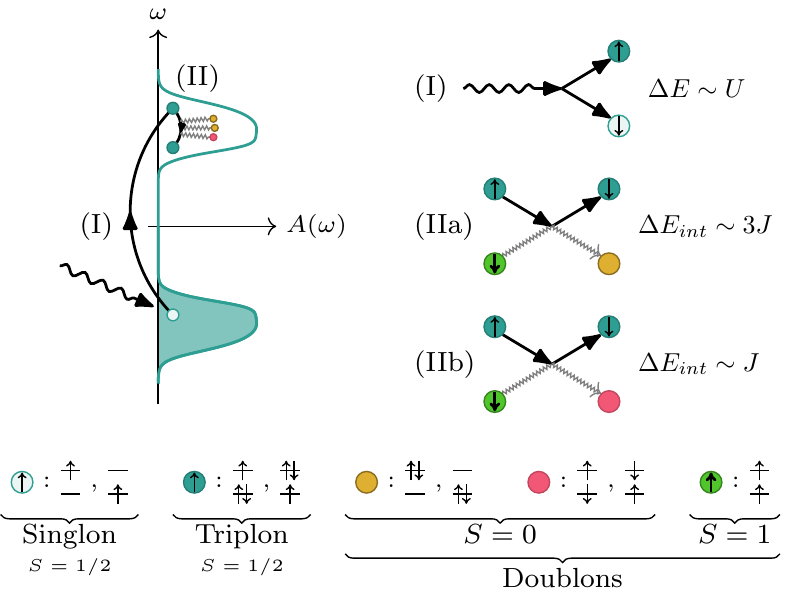}\\[-3mm]
  \caption{\label{fig:Schematic}(Color online) Schematic illustration of the photo-doping process in a two-band Mott insulator, creating photo-excited singlon-triplon pairs (I), and the subsequent relaxation by doublon-triplon (and doublon-singlon) scattering (II), which generates high-energy (IIa) and low-energy (IIb) $S=0$ doublon excitations on the $S=1$ high-spin doublon background. A legend for the local two-orbital states is shown at the bottom.
    %
  }
\end{figure}
%


In multiorbital Hubbard systems one should expect similar processes to play a prominent role even in the absence of phonon couplings, antiferromagnetic correlations, and nonlocal interactions.
In Mott insulators with small or vanishing crystal field splitting, the Hund's coupling controls the energies of the most relevant local states, and favors high-spin states.
The goal of this study is to clarify how photo-doped carriers moving in this high-spin background produce local spin excitations and thereby transfer their excess kinetic energy to interaction energy. 

We will explore this physics in the simple context of a half-filled two-orbital Mott insulator with degenerate bands and density-density interactions. For positive Hund's coupling $J$, the dominant states are the half-filled high spin states ($[\uparrow, \uparrow]$) indicated by the green dots in Fig.~\ref{fig:Schematic}, while the two low-spin states ($[\uparrow,\downarrow]$ and $[\uparrow\downarrow,0]$) have an energy that is higher by $J$ (red dots) and $3J$ (yellow dots), respectively.
Driving the system with a frequency above the inter-orbital interaction $U$ produces photo-carriers in terms of singlon-triplon pairs, as illustrated by process (I) of Fig.~\ref{fig:Schematic}. The singlons and triplons, with an initially high kinetic energy, then scatter with the high-spin doublon states transferring energy quanta of $J$ (process IIa) or $3J$ (process IIb) to local spin excitations.
As a result of this, the energy distributions of the triplons changes, and the corresponding population shifts within the upper Hubbard band towards the lower band edge. (An analogous dynamics takes place in the lower Hubbard band, where the singlon distribution shifts to the upper band edge.)
We also observe a splitting of the photo-carrier distribution in the upper Hubbard band, which can be linked to the Hund's coupling induced splitting of the local states within the doublon sector. By systematically studying the relaxation at a constant $1\%$ photo-doping as a function of temperature and Hund's coupling we furthermore demonstrate two mechanisms, \emph{thermal blocking} and \emph{kinetic freezing} that partially inhibit the relaxation dynamics within the doublon sector.


This paper is organized as follows.  
Section~\ref{sec:model} presents the effective two-band Hubbard model, and Sec.\ \ref{sec:method} presents the real-time dynamical mean-field method that we use to study the non-equilibrium dynamics.
In Sec. \ref{sec:results} we present the equilibrium spectral function (Sec.\ \ref{sec:equilibrium}) and then discuss the photo doping dynamics (Sec.\ \ref{sec:photodoping}).
The different aspects of the relaxation dynamics are explored in the
following sections:
Section \ \ref{sec:tdpes} presents the simulated time-resolved photo emission spectra, while
Sec. \ref{sec:shorttimeExcitations} analyzes the short time dynamics. 
The doublon excitation dynamics  is discussed in Sec.\ \ref{sec:doublonExcitations},
and its temperature and Hund's coupling dependence  in Secs.\ \ref{sec:Tsweep} and \ref{sec:Jsweep}, respectively.
Section \ref{sec:discussionAndConclusion} contains a discussion and conclusions.

\section{Model}
\label{sec:model}

The two-band Hubbard model is the canonical model for interacting multiband electron systems, relevant e.g.\ for the $e_{g}$ irreducible representation of d-orbitals in crystal fields with cubic or tetragonal symmetry \cite{Sugaon:1970kx}. The Hamiltonian has the form
\begin{equation}
 \hat{H} = 
 -t_\text{hop} \sum_{\langle i, j \rangle} \sum_{\alpha \sigma}
 (\cc_{i \alpha \sigma} \ca_{j \alpha \sigma} + \cc_{j \alpha \sigma} \ca_{i \alpha \sigma})
 + \sum_{i} \hat{H}_{\text{loc}, i}
 \, ,
  \label{eq:Hamiltonian}
\end{equation}
where $\cc_{j\alpha\sigma}$ creates an electron on site $i$ in orbital $\alpha=1$, $2$ with spin $\sigma = \uparrow$, $\downarrow$, $t_\text{hop}$ is the nearest-neighbor hopping, 
and $\hat{H}_{\text{loc}, i}$ is the local interaction Hamiltonian on site $i$. The interaction $\hat{H}_{\text{loc}, i}$ has the general structure
\begin{multline}
 \hat{H}_{\text{loc}, i} = 
 U \sum_{\alpha} \n_{i \alpha \uparrow} \n_{i \alpha \downarrow}
 + \sum_{\alpha \ne \beta} \sum_{\sigma, \sigma'}
   (U' - J\delta_{\sigma \sigma'}) \n_{i \alpha \sigma} \n_{i \beta \sigma'} \\
 + \gamma J \sum_{\alpha \ne \beta} \left(
 \cc_{i \alpha \uparrow} \cc_{i \alpha \downarrow} \ca_{i \beta \downarrow} \ca_{i \beta \uparrow}
 +
 \cc_{i \alpha \uparrow} \cc_{i \beta \downarrow} \ca_{i \alpha \downarrow} \ca_{i \beta \uparrow}
 \right)
 \, ,
 \label{eq:HlocHund}
\end{multline}
where $U$ is the intra-orbital Coulomb interaction, $U'$ is the inter-orbital Coulomb interaction, and $J$ is the Hund's coupling. The interaction in Eq.~(\ref{eq:HlocHund}) is the well known Kanamori interaction \cite{Kanamori:1963aa}, which becomes rotationally invariant when $U' = U - 2J$ and $\gamma = 1$.

\indent Here we set $U' = U - 2J$ but employ the non-symmetric form (with $\gamma = 0$) for practical reasons. This approximation, which only retains the density-density interactions, has been shown to have a richer phase diagram than the isotropic model \cite{Hoshino:2016aa, Steiner:2016aa}.
The lattice is assumed to be bipartite, and the non-interacting dispersion is chosen to yield a semi-circular density of states with bandwidth $W = 4t_\text{hop}$. In the following, we use $t_\text{hop}$ as the unit of energy, so that time is measured in units of $\hbar/t_\text{hop}$. The chemical potential is fixed to $\mu = (3U - 5J)/2$ which yields half-filling (two electrons per site).
The system undergoes an antiferromagnetic spontaneous symmetry breaking at low temperature \cite{Hoshino:2016aa, Steiner:2016aa}, but in the present study we restrict our attention to the paramagnetic high-temperature Mott phase.

\section{Method}
\label{sec:method}

To investigate the non-equilibrium dynamics of the two-band Hubbard model [Eq.\ (\ref{eq:Hamiltonian})] in the strongly correlated paramagnetic Mott insulating phase, we employ real-time dynamical mean-field theory
(DMFT) \cite{Aoki:2014kx}. The DMFT formalism neglects the momentum dependence of the self-energy, i.e.\ $\Sigma(\mbf{k}, t; \mathbf{k}', t') \approx \Sigma(t, t')$, which allows to map Eq.~(\ref{eq:Hamiltonian}) to an impurity action
\begin{equation}
  S =
  \int_\mathcal{C} dt \hat{H}_{loc}(t) +
  \iint_\mathcal{C} dt dt'
  \sum_{\alpha \sigma}
  c^\dagger_{\alpha\sigma}(t) \Delta_{\alpha \sigma} (t, t') c_{\alpha \sigma}(t')
  \label{eq:ImpurityAction}
\end{equation}
with a self-consistently determined dynamical mean-field in the form of a two-time-dependent hybridization function $\Delta_{\alpha \sigma}(t, t')$.
The featureless semi-circular density of states allows us to capture the general physics of the Mott insulator, and yields the simple self-consistency relation $\Delta_{\alpha \sigma} = t_\text{hop}^2 g_{\alpha \sigma}$, where $g_{\alpha \sigma}$ is the impurity single-particle Green's function, $g_{\alpha\sigma}(t, t') \equiv -i \langle \mathcal{T} c_{\alpha \sigma}(t) c^\dagger_{\alpha \sigma}(t') \rangle_S$.

Since the Mott insulator is out-of-reach for weak coupling DMFT approaches \cite{Tsuji:2013mi, Tsuji:2013kkk} and variational approaches \cite{Oelsen:2011cr, Oelsen:2011nx, Behrmann:2013aa, Behrmann:2015aa, Sandri:2015aa, Behrmann:2016aa, Mazza:2017aa}, we employ a pseudo particle strong coupling (PPSC) real-time impurity solver. Introducing pseudo particles enables diagrammatic expansions in the hybridization function, and we employ the first and second order self-consistent dressed approximation, also known as the non-crossing approximation (NCA) and one-crossing approximation (OCA) \cite{Eckstein:2010fk}. Since the dressed self-energy approximation can be written as a functional derivative of a (pseudo particle) Luttinger-Ward functional \cite{Baym:1961tw, Baym:1962qo}, the PPSC expansion yields conserving approximations for both density and energy. These conservation laws are central for studying the time evolution of the system, especially at longer times. A detailed description of the method can be found in Ref.~\onlinecite{Eckstein:2010fk}.

%
We have recently extended our real-time PPSC solver to general multi-orbital systems, using both the blocking of the local Hamiltonian and a symmetry analysis of the pseudo particle diagrams to reduce the computational effort. The numerical symmetry analysis is flexible enough to automatically identify special situations such as spin and particle-hole symmetry, which for the paramagnetic half-filled two-band Hubbard model [Eq.\ (\ref{eq:Hamiltonian})] with density-density interactions reduces the number of pseudo-particle NCA self-energy diagrams from 64 to 10, and the number of NCA single-particle Green's function diagrams from 32 to 12.

Real-time DMFT directly gives the local single-particle Green's function $g_{\alpha\sigma}(t, t')$ of the system and the local many-body density matrix $\rho(t)$ through the pseudo-particle Green's function $\hat{G}(t, t')$, $\rho(t) = -i\hat{G}^<(t, t)$. In the case of density-density interaction both $\hat{G}$ and $\rho$ are diagonal in the local occupation number basis $\Gamma$, i.e.\ $\rho_{\Gamma \Gamma'} = \delta_{\Gamma\Gamma'} \rho_{\Gamma}$.
%
%
Note that the trace over the local many-body density matrix is a conserved quantity, i.e.\
\begin{equation}
  \sum_\Gamma \rho_{\Gamma}(t) = 1
  \, \Rightarrow \,
  \sum_\Gamma \partial_t \rho_{\Gamma}(t) = 0
  \, .
  \label{eq:ProbabilityDensity}
\end{equation}
The total energy of the system $E_\text{tot}$ is the sum of the interaction ($E_\text{int}$) and kinetic ($E_\text{kin}$) energy, $E_\text{tot} = E_\text{int} + E_\text{kin}$. Using the local multiplet energies $E_{\Gamma}$, $E_\text{int}$ is obtained as
\begin{equation}
  E_\text{int}(t) = \sum_\Gamma \rho_\Gamma(t) E_\Gamma
  \, ,
  \label{eq:Eint}
\end{equation}
and $E_\text{kin}$ is given by the equal-time contour convolution between $\Delta$ and $g$,
$E_\text{kin}(t) = -i \sum_{\alpha\sigma} [\Delta_\alpha \ast g_{\alpha\sigma}]^<(t, t)$,
see Ref.\ \onlinecite{Aoki:2014kx}.

The time evolution is solved using a fifth-order multi-step method with backwards differentiation, and equidistant steps $\Delta t = 0.01 t_\text{hop}$ in real time and $N_\tau = 100$ to $800$ steps in imaginary time $\tau$, $\tau \in [0,\beta)$, where $\beta$ is the inverse temperature $\beta = 1/T$. The discretization errors are controlled by ensuring that the drift in the total density per site is below $10^{-4}$. We note that the density conservation is sensitive to both real and imaginary time discretizations and therefore gives a good measure of the total convergence.

\section{Results}
\label{sec:results}

\subsection{Equilibrium}
\label{sec:equilibrium}

The equilibrium properties of the Mott phase in the strongly interacting half-filled two-band Hubbard model [Eq.\ (\ref{eq:Hamiltonian})] are well understood within DMFT \cite{Koga:2005aa, Pruschke:2005aa, Werner:2006qy, Sakai:2006aa, Werner:2007lr, Bulla:2008kx, Peters:2010aa, Peters:2011aa, Antipov:2012aa}.
In what follows we fix the interaction to $U = 15$ and the band width to $W=4$, which corresponds to the deep Mott insulator regime, $U \gg W$. 
In equilibrium, a numerically exact DMFT solution can be obtained using the continuous time quantum Monte Carlo (CT-QMC) method \cite{Werner:2006rt,Werner:2006qy,Gull:2011lr}, which allows to gauge the accuracy of our PPSC approach in the Mott regime.

%
%
\begin{figure}
  \includegraphics[scale=1]
  {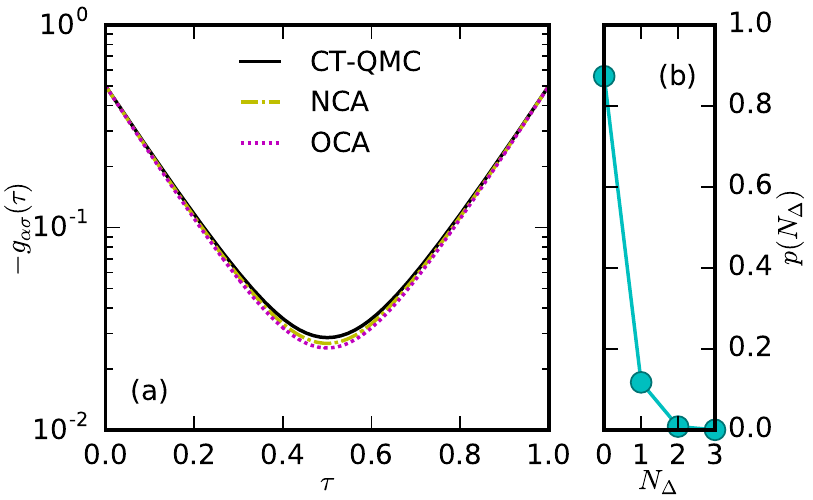}
  \caption{\label{fig:Gtau}(Color online) Panel a: Equilibrium imaginary time single particle Green's function $g_{\alpha \sigma}(\tau)$ for $\beta=1$, $U=15$, and $J/U=0.06$. Comparing exact continuous-time quantum Monte Carlo (CT-QMC), with first (NCA) and second order (OCA) strong coupling expansions.
    Panel b: CT-QMC hybridization expansion probability contribution $p(N_\Delta)$ to the partition function ($N_\Delta$ is the perturbation order).
  }
\end{figure}
%

In Fig.\ \ref{fig:Gtau}a we compare the equilibrium CT-QMC solution for the local single-particle Green's function $g_{\alpha \sigma}(\tau)$ in imaginary time $\tau$, obtained using the hybridization expansion as implemented in the Triqs application cthyb \cite{Seth2016274,Parcollet2015398,Boehnke:2011fk}, with the NCA and OCA result at inverse temperature $\beta=1$ and the relative Hund's coupling $J/U = 0.06$.
We see that the strong local interaction ($U=15$) makes the first order strong coupling approximation (NCA) an extremely good approximation. The deviation between the Green's function obtained from NCA/OCA and CT-QMC are lower than the stochastic QMC noise $|g_{\text{NCA/OCA}}(\tau) - g_{\text{QMC}}(\tau)| < 0.004$ for $\beta=1$ (and $\beta=10$, not shown).
Comparing NCA and OCA we find that in contrast to the metallic single band case \cite{Eckstein:2010fk} the OCA result is more correlated than the NCA result, i.e., $g_{\text{OCA}}(\beta/2) < g_{\text{NCA}}(\beta/2)$.

The good performance of NCA can be understood by looking at the distribution over hybridization perturbation orders in the CT-QMC solution, see Fig.\ \ref{fig:Gtau}b. In the deep Mott state the expansion of the partition function is dominated by the zeroth order perturbation order $N_\Delta = 0$ (the atomic limit) with a $\sim 10\%$ contribution from the first order terms $p(N_\Delta \! = \! 1) \approx 0.1$, corresponding to a single hybridization insertion, while the second order terms contribute less than one percent $p(N_\delta \! = \! 2) \lesssim 0.008$. The zeroth and first order contributions are exactly captured within NCA, and the higher order contributions are exponentially small, hence the close agreement with CT-QMC.

%
%
\begin{figure}
  \includegraphics[scale=1]
  {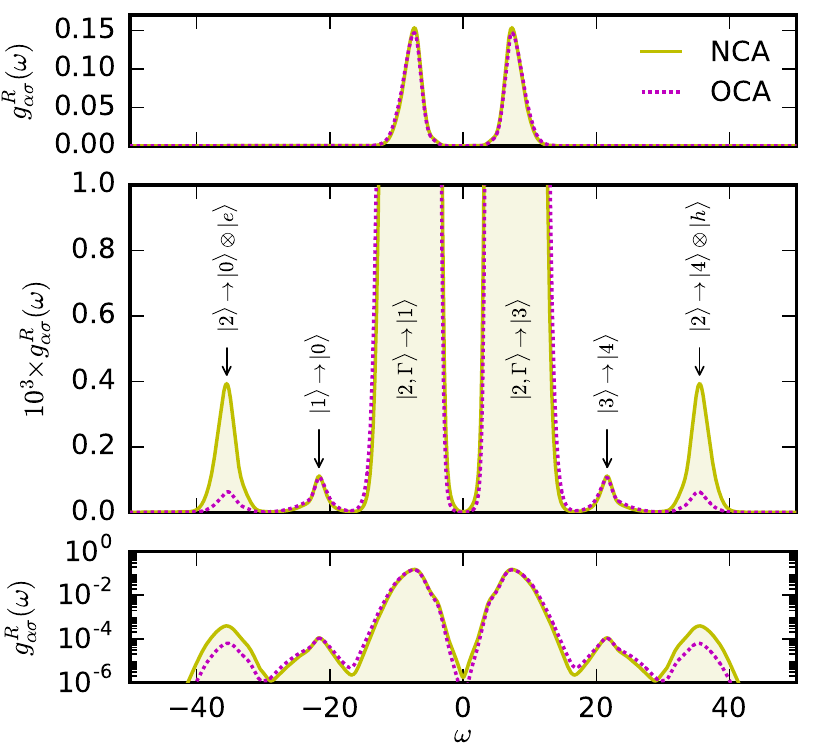}
  \caption{\label{fig:Gw}(Color online) Real-frequency equilibrium spectral function for $\beta=1$, $U=15$, and $J/U = 0.06$.
  }
\end{figure}
%
%
By time-evolving the equilibrium solution within NCA and OCA we can directly obtain the real-frequency single-particle spectral function by Fourier transforming the real-time Green's function, see Fig.\ \ref{fig:Gw}. The dominant spectral features of the half-filled Mott insulator are the lower and upper Hubbard bands centered around $\omega \approx \pm U/2 = \pm 7.5$. The single particle spectral function corresponds to particle addition and particle removal for positive and negative frequencies $\omega$ respectively. Thus the upper (lower) Hubbard band corresponds to the particle addition (removal) processes $|2, \Gamma \rangle \rightarrow |3 \rangle$ ($|1\rangle$), where $\Gamma$ is one of the doublon states listed in Fig.\ \ref{fig:Schematic}.
In contrast to the single band Hubbard model the larger Hilbert space of the two-band model results in two additional spectral features at energies beyond the Hubbard bands. These resonances come in two classes, thermally activated and quantum fluctuation driven, analogous to what has previously been discussed for the Bose-Hubbard model \cite{Strand:2015ac}.
The first resonance beyond the Hubbard bands arises from thermally activated electron-hole pair excitations in the ground state, which on particle addition (removal) produces the local transitions $|3 \rangle \rightarrow |4\rangle$ ($|1 \rangle \rightarrow |0 \rangle$).
While these resonances involve only the multiplet states at one site, the second resonance corresponds to concomitant addition (removal) of an electron and the emission of a hole $|h\rangle$ (particle $|p\rangle$) excitation into other lattice sites, i.e.\ $|2 , \Gamma\rangle \rightarrow |4\rangle \otimes |h\rangle$ ($|2 , \Gamma\rangle \rightarrow |0\rangle \otimes |p\rangle$).
As seen in Fig.\ \ref{fig:Gw}, NCA overestimates the spectral weight of the second resonance in comparison to OCA. However, since this resonance is very high in energy its influence on the photodoping dynamics across the Hubbard gap is negligible. Hence, we will apply NCA in the study of the photo-doping dynamics of the deep Mott insulator in what follows.

\subsection{Non-equilibrium photo doping}
\label{sec:photodoping}

To study the response of the two-band Mott insulator relevant to pump-probe experiments \cite{Aubock:12, Baldini:2016aa, Baldini:2017aa}, we use a simplified driving to produce particle-hole transitions across the Mott gap (process I in Fig.\ \ref{fig:Schematic}).
Specifically, we employ a two-cycle modulation of the single particle hopping with frequency $\Omega_\text{pump}= 2.25U$ and a Gaussian envelope. 
This mechanism is different from the application of an electric field, but since we aim at investigating the relaxation {\em after} the excitation, the precise mechanism of generating carriers is not important, as long as it produces an essentially instantaneous excitation of particle-hole pairs across the Mott gap.

%
We focus again on the strongly correlated Mott insulator with Hubbard $U=15$ and a single particle bandwidth fixed to $W=4$.
The recombination rate of triplons and singlons is in this case exponentially suppressed by the large Mott gap ($U \gg W$) \cite{Sensarma:2010aa, Lenarifmmode-celse-cfiiifmmode-celse-cfi:2014aa, Lenarifmmode-celse-cfiiifmmode-celse-cfi:2013aa} and the photo-doping generated by the pump-excitation is therefore an approximately conserved quantity on the numerically accessible timescales. In addition, to enable a direct comparison between systems with different strengths of the Hund's coupling $J$, we fix the density of photo-doping induced singlon-triplon carriers after the pump to $1\%$ by tuning the pump amplitude.

The system responds to the pump-excitation on all the timescales corresponding to characteristic energies of the Hamiltonian [Eq.\ (\ref{eq:Hamiltonian})], i.e., i) the Hubbard $U=15$, ii) the single particle hopping bandwidth $W=4$, and iii) the Hund's coupling $J \lesssim 0.05 U = 0.75$.
However, before we disentangle the photo-doping dynamics in terms of the kinetic energy and the local many-body state occupations, we present results for the time-dependent photo emission spectra.

\subsubsection{Time-dependent photo-emission spectroscopy}
\label{sec:tdpes}

%
%
\begin{figure*}
  \includegraphics[scale=1]
  {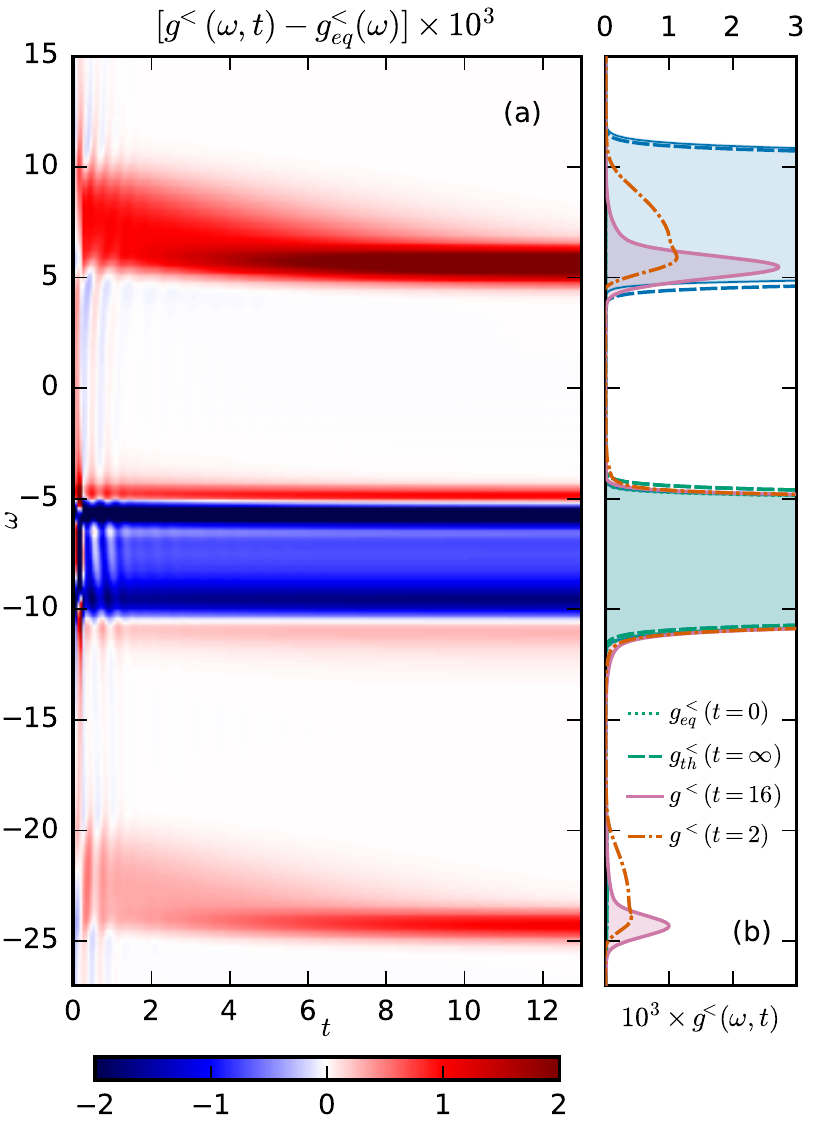}
  \ \ \ \
  \includegraphics[scale=1]
  {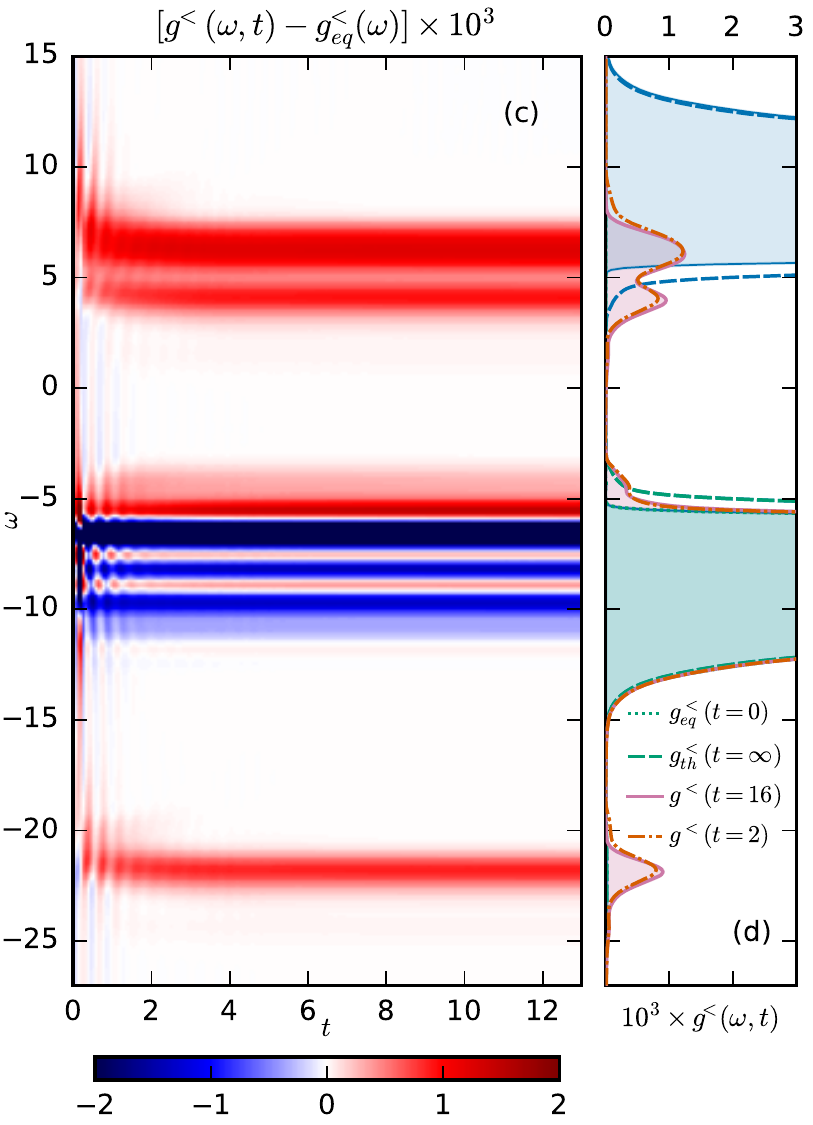}
  \caption{\label{fig:tdpes}(Color online) Theoretical time dependent photo-emission spectra showing the kinetic energy relaxation of the photo-doped triplons in the upper Hubbard band for $U=15$, $T=0.1$, $J/U=0.01$ (panels a \& b), and $J/U=0.06$ (panels c \& d).}
\end{figure*}
%

%
%
\begin{figure*}
  \includegraphics[scale=1]
  {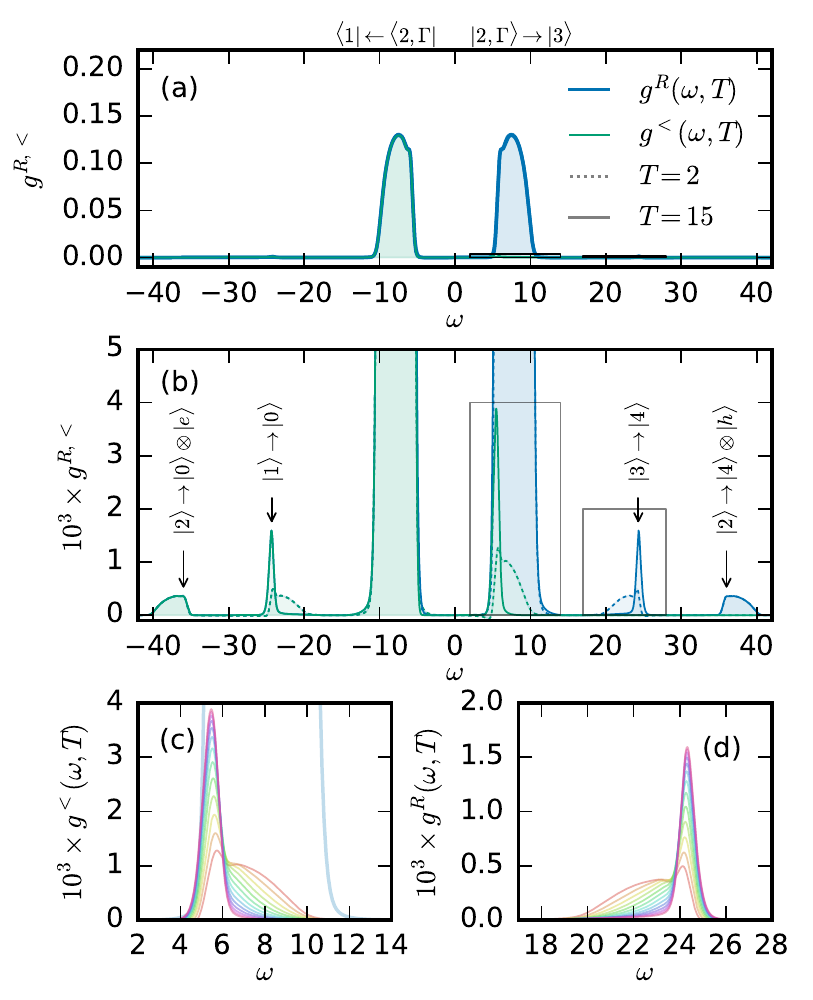}
  \includegraphics[scale=1]
  {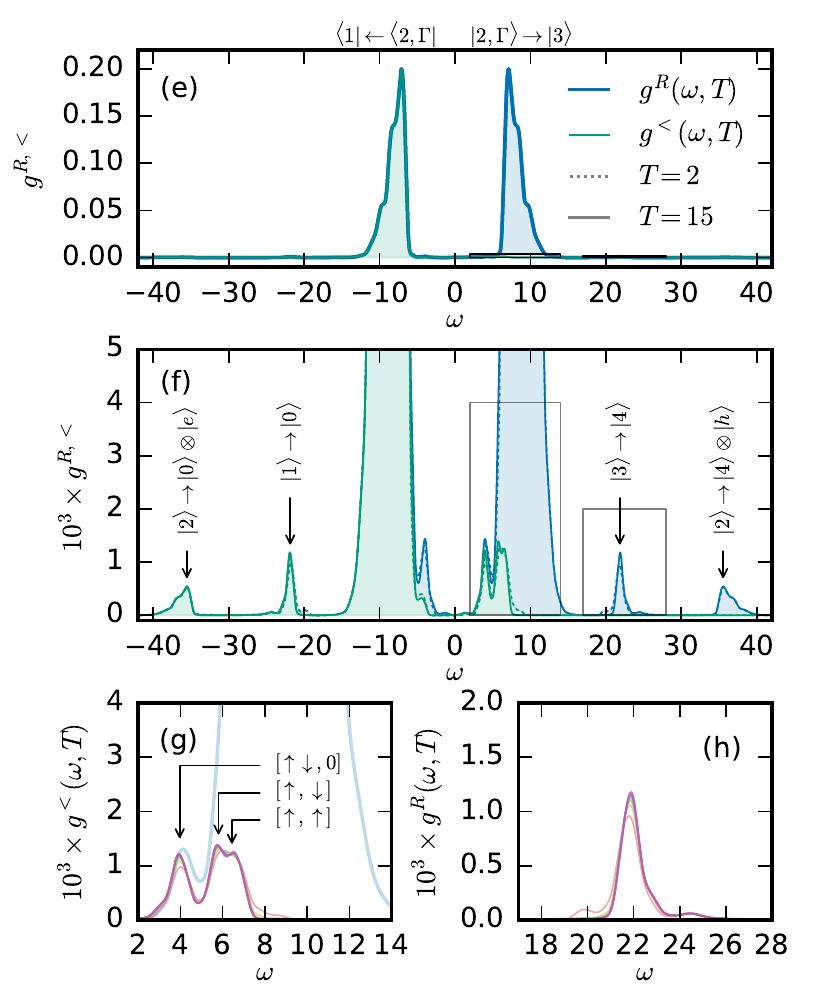} \\[-5mm]
  \caption{\label{fig:spA}(Color online) Time-evolution of the single-particle retarded Green's function $g^R(\omega,T)$ (blue) and lesser Green's function $g^<(\omega, T)$ (green), for $U=15$, $\beta=10$, $J/U=0.01$ (panels a-d), and $J/U= 0.06$ (panels e-h). Panels c,d,g and h show all integer times $t=2$ to $14$ (red to blue).}
\end{figure*}
%

To show how multi-orbital and Hund's physics may be detected experimentally using time-dependent photo-emission spectroscopy (TD-PES) 
\cite{Rameau:2016aa, Kemper:2016aa}, we study the time evolution of the non-equilibrium single-particle spectral function after the pump. Disregarding matrix elements and the finite width of the probe pulse this corresponds to the theoretical time-dependent photo-emission spectrum \cite{Freericks:2009aa}.
The photo-doping generates a redistribution of the spectral weight in both the retarded spectral function $g^R(\omega, t)$ and the lesser spectral function $g^<(\omega, t)$ defined as
\begin{equation}
  g^{\gamma}(\omega, t) = \pm\frac{1}{\pi} \text{Im}
  \int_t^{\infty} d\bar{t} \, 
  e^{-i (\bar{t}-t) \omega}  
  g^\gamma( t, \bar{t} )
  \, ,
\end{equation}
where $\omega$ is the relative frequency, $t$ is the absolute time, the $+$($-$) sign corresponds to the lesser, $\gamma = <$ (retarded, $\gamma=R$) component.

Since the lesser spectral function $g^<(\omega, t)$ -- i.e.\ the occupied density of states (DOS) -- is the main observable in TD-PES, we focus on the time-dependent change in $g^<(\omega, t)$ relative to the time-translational invariant occupied DOS $g_\text{eq}^<(\omega)$ of the initial equilibrium state. In Fig.\ \ref{fig:tdpes} the differential response of the occupied DOS is shown for $J/U=0.01$ (Fig.\ \ref{fig:tdpes}a-b) and $J/U=0.06$ (Fig.\ \ref{fig:tdpes}c-d).

The short pulse injects particle-hole pairs (singlon-triplon pairs) with high kinetic energy into the Hubbard bands (process I in Fig.\ \ref{fig:Schematic}), and thus results in an initial spectral weight redistribution from the lower to the upper Hubbard band. The singlons and triplons then relax their high kinetic energy by emitting doublon spin-excitations (processes IIa and IIb in Fig.\ \ref{fig:Schematic}) when scattering off the high spin doublon background, as indicated in the upper Hubbard band in Fig.~\ref{fig:Schematic}.
The result is a drift of the occupied density of states in the upper Hubbard band from the upper band edge to the lower band edge. 
However, as can be seen by comparing Fig.\ \ref{fig:tdpes}a-b with \ref{fig:tdpes}c-d the relaxation dynamics and the spectral features induced by this cooling process depend strongly on the Hund's coupling $J$.

At small Hund's coupling, e.g.\ $J/U=0.01$  as in Fig.~\ref{fig:tdpes}a-b, the kinetic relaxation of triplons manifests itself primarily as a redistribution of occupied spectral weight within the upper Hubbard band. A short time after the pulse (see $t=2$ dot-dashed curve in Fig.\ \ref{fig:tdpes}b) the broad distribution in the upper Hubbard band indicates the presence of triplons with high kinetic energy. This population relaxes at later times to a thermal-like distribution (solid red line), albeit with a shifted chemical potential. While the triplons apparently ``thermalize'' within the upper Hubbard band, the large Mott gap exponentially suppresses singlon-triplon recombination, which is required for the system to reach a true thermal equilibrium state.

The observed population dynamics within the Hubbard band (Fig.\ \ref{fig:tdpes}a-b) is similar to the dynamics of the photo-doped anti-ferromagnetic systems \cite{Goleifmmode-zelse-zfi:2014aa, Eckstein:2016aa}
or systems coupled to the lattice degrees of freedom \cite{0295-5075-109-3-37002,Werner:2013uq,Goleifmmode-zelse-zfi:2012aa}, where the electrons interact with external bosonic degrees of freedom.
Thus, at weak Hund's coupling the local doublon spin-degrees of freedom in the two band model qualitatively act as a boson-bath into which the electronic degrees of freedom can transfer excess kinetic energy.

We also note that the dynamics in the Hubbard band is reproduced in the second Hubbard resonance at $\omega \approx -3U/2 = -22.5$, see Fig.\ \ref{fig:tdpes}a-b.
This spectral feature corresponds to the particle removal transition $| 1 \rangle \rightarrow | 0 \rangle$, as previously discussed in Fig.\ \ref{fig:Gw}.
Note that the singlon population (corresponding to the unoccupied DOS)  in the lower Hubbard band is the mirror copy of the triplon population in the upper Hubbard band, by particle hole-symmetry (not shown).
Hence, while triplons relax towards lower frequencies at the lower band edge of the upper Hubbard band, singlons relax towards higher (small negative) frequencies at the upper band edge of the lower Hubbard band. In contrast, the resonance $| 1 \rangle \rightarrow | 0 \rangle$ at $\omega \approx -22.5$ in the occupied density of state represents photo-emission from a singlon state, and displays a reversed relaxation towards low (large negative) frequencies.
Since the multiplet energy separation between the singlon and holon state is equal to $U$ the singlons at the upper band edge of the lower Hubbard band with large (small negative) kinetic energy require a smaller excitation energy in the $|1 \rangle \rightarrow |0 \rangle$ transition, whence these singlons produce the low frequency structure in the $\omega = -3U/2$ resonance. 

While this spectral feature is present in equilibrium at elevated temperatures due to thermal activation of singlon states [Fig.\ \ref{fig:Gw}], its emergence after the pump is a non-equilibrium effect induced by the photo-doped singlons.
We note that the emergence of satellites below the lower Hubbard band is a generic feature of multiorbital systems with Hubbard $U$ interaction, due to the additional states in the local Fock space with higher/lower occupation numbers.
The generation of this class of side bands in photo-doped multi-orbital could be used to experimentally detect the presence of multiorbital interactions in an out-of-equilibrium set-up.

For larger $J/U$ the time scale of the kinetic relaxation becomes shorter, and for $J/U=0.06$ the initial high to low kinetic energy triplon redistribution overlaps with the short lived coherent $2\pi/U$-oscillations at $t \lesssim 2$, see Fig.~\ref{fig:tdpes}c-d.
In this regime the relaxation changes its qualitative behavior. After the fast kinetic relaxation, the occupied spectral weight in the upper Hubbard band no longer comprises a single resonance, instead, the spectral weight is split into two main resonances (at $\omega \approx 4$ and $\omega \approx 7$), see Fig.~\ref{fig:tdpes}d. Furthermore, the second Hubbard-resonance (at $\omega=-22$) is no longer a mirror copy of the entire occupied spectral weight in the upper Hubbard-band, but rather only a mirror copy of the low energy resonance.

The splitting of the occupied spectral weight in the upper Hubbard band of Fig.\ \ref{fig:tdpes}c-d can be understood in terms of the Hund's coupling induced splitting of the doublon states, where the $S=0$ doublon states $[\uparrow, \downarrow]$ and $[\uparrow\downarrow,0]$ (see Fig.\ \ref{fig:Schematic}) are split off from the high-spin $S=1$ doublon states $[\uparrow,\uparrow]$ by $J$ and $3J$ respectively.
After the pump excitation the excess triplon density can transition to all three doublon states on particle removal, which directly produces three resonances in the upper Hubbard band, corresponding to the processes $|3\rangle \rightarrow |2,\Gamma\rangle$, see also the discussion of Fig.\ \ref{fig:spA}g below. 
Note that the particle hole symmetric process $|1\rangle \rightarrow |2, \Gamma \rangle$ produces a shoulder feature at the upper edge of the lower Hubbard band, at $\omega \approx -5$ in Fig.\ \ref{fig:tdpes}d, which might be simpler to detect experimentally than the occupied DOS in the upper Hubbard band.
We propose that the detection of a splitting of the photo-doped occupation in the upper Hubbard band and the emergence of a low energy shoulder of the lower Hubbard band can serve as a  ``litmus test'' for strong Hund's coupling in Mott insulators. 

A similar reasoning also explains why the second-Hubbard resonance (at $\omega \approx -23$) only displays a single peak, see Fig. \ref{fig:tdpes}c-d. This spectral feature corresponds to the removal of 
a photo-doping induced singlon state, i.e.\ $|1\rangle \rightarrow |0\rangle$. Since these states are not split by the Hund's coupling $J$ the transition only yields a single peak.

%

So far we have focused on the occupied density of states since it is experimentally observable in TD-PES. However, our real-time DMFT calculations give direct access also to the retarded spectral function $g^R(\omega, t)$. Thus, we are able to give the complete picture of the spectral distribution throughout the relaxation, see Fig. \ref{fig:spA}.
Beyond the features already identified in the differential TD-PES, the full spectral function shows the general shape of the Hubbard bands at $J/U=0.01$ and $0.06$, see Fig.~\ref{fig:spA}a and \ref{fig:spA}e. Going from low to high $J$ the Hubbard bands develop from a resonance with a low frequency shoulder, into a resonance with two shoulders on the high frequency side.
In addition, panels Fig.\ \ref{fig:spA}b and \ref{fig:spA}f show not only the redistribution due to kinetic relaxation but also that the third resonances beyond the upper and lower Hubbard bands (at $|\omega| \approx 36$) are not affected by the pump excitation. This can be understood by considering the local transitions that give rise to the resonance, e.g.\ at positive frequencies they correspond to $|2, \Gamma \rangle \rightarrow |4\rangle \otimes |h\rangle$ (where $|h \rangle$ is an emitted hole on the lattice). Since the singlon-triplon states are not directly involved in these transitions, their relaxation does not affect this third class of resonances. 

Detailed plots of time slices at the upper Hubbard band and the second resonance clearly show that at low $J/U = 0.01$ they are mirror copies of each other, see Fig.\ \ref{fig:spA}c and \ref{fig:spA}d. At this low Hund's coupling the thermal activation in the initial state with $T=0.1$ makes it impossible to resolve the individual doublon excitations since their splitting $J=0.15$ is of the same order, $J \approx T$.
At larger Hund's coupling $J/U=0.06$ however, the splitting of the atomic multiplets gives rise to a separation of the occupied DOS in the upper Hubbard band, see Fig.~\ref{fig:spA}f and \ref{fig:spA}g. In this particular case all three resonances are resolved in Fig.~\ref{fig:spA}f (due to better spectral resolution as compared to Fig.\ \ref{fig:tdpes}d).
We also note that the shoulder feature in the lower Hubbard band is even more pronounced in the retarded spectral function at $\omega \approx -5$ as compared to the feature in the occupied DOS, c.f.\ green and blue lines in Fig.\ \ref{fig:spA}f.
Further, while the second resonance qualitatively is a single peak [see Fig.\ \ref{fig:spA}h], a close inspection reveals additional structures. In particular it has a high energy extended shoulder feature due to higher order processes.

\subsubsection{Short time particle-hole excitations}
\label{sec:shorttimeExcitations}
%
In order to understand the photo-carrier relaxation observed in our time-dependent PES results, in particular its dependence on the relative Hund's coupling $J/U$ and temperature $T$, we perform a detailed analysis of the relaxation dynamics. To do this we will exploit the fact that our PPSC method also provides information about the local many-body density matrix $\rho(t)$, and use this to study the time-evolution of the different local many-body states throughout the relaxation process as a function of $T$ and $J/U$.

%
%
\begin{figure}
  \includegraphics[scale=1]
  {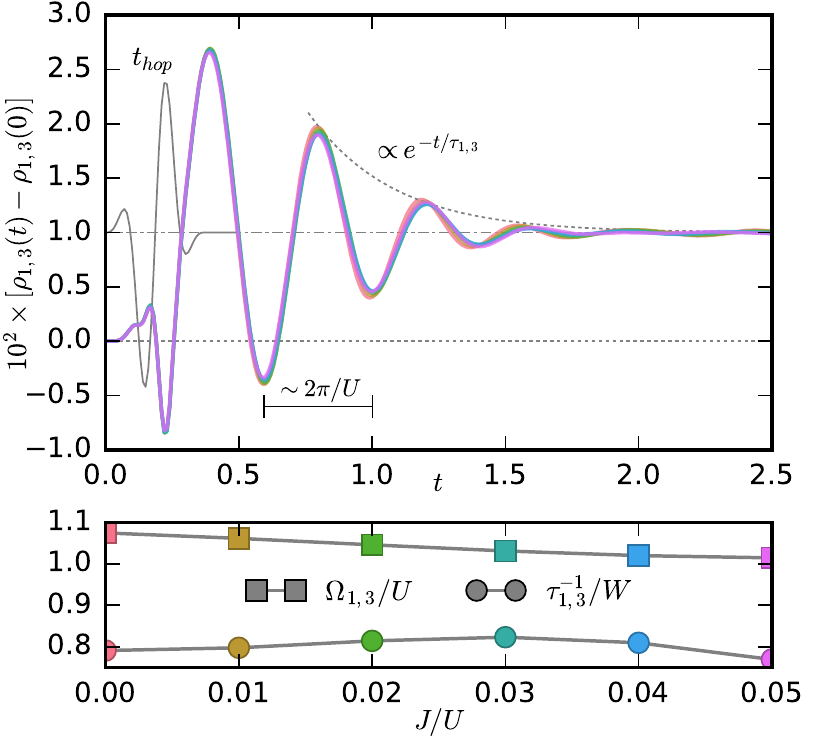}\\[-4mm]
  \caption{\label{fig:DopingDamping}(Color online) Short-time dependence of the photo-doping pump-induced change in the singlon-triplon density $\rho_{1,3}(t) - \rho_{1,3}(0)$ for $U=15$, $\beta=1$, and $J/U=0.00$ to $0.05$ (upper panel), showing damped $2\pi/U$ oscillations $\Omega_{1,3} \sim U$ with an inverse relaxation time $\tau^{-1}_{1,3}$ of the order of the 
  bandwidth $W$ 
  (lower panel). The hopping modulation $t_\text{hop}(t)$ of the pump pulse is shown as a gray thin solid line in the upper panel.}
\end{figure}
%
%
%
%
\begin{figure}
  \includegraphics[scale=1]
  {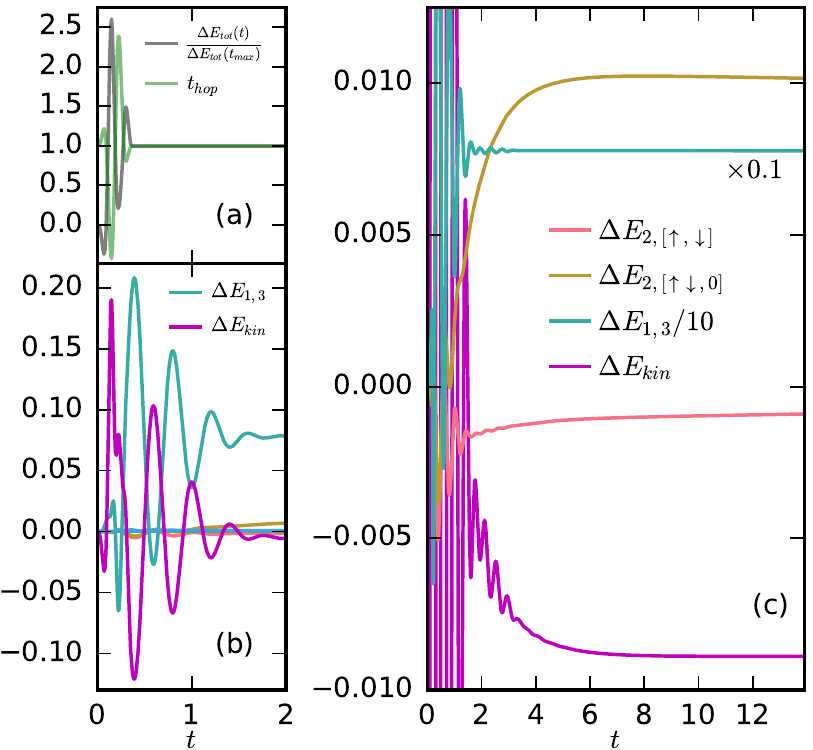}
  \caption{\label{fig:EnergyConversion}(Color online) Three stages of excitation/relaxation: (a) Hopping modulation $t_\text{hop}$ (green) and increase of the total energy $E_\text{tot}$ (black) at $t \lesssim 0.5$, (b) energy conversion between kinetic energy $E_\text{kin}$ (magenta) and singlon-triplon interaction energy $E_{1,3}$ (blue) at $t \lesssim 2$ (c.f.\ Fig.\ \ref{fig:DopingDamping}), and (c) conversion from (singlon-triplon) kinetic energy to interaction energy in terms of $S=0$ doublon excitations $E_{2, [\uparrow, \downarrow]}$ (red) and $E_{2, [\uparrow\downarrow, 0]}$ (yellow), c.f. (IIa) and (IIb) in Fig.\ \ref{fig:Schematic}. The approximately conserved singlon-triplon energy $\Delta E_{1,3}$ (rescaled by 0.1) is also shown (blue).}
\end{figure}
%
%
The short time response of the system at $t \lesssim 2$, shown in Fig.\ \ref{fig:DopingDamping}, displays only the two fastest time scales $U$ and $W$. The response during the application of the pump (gray line) is dominated by a rapid increase of the singlon and triplon probability $\rho_{1}=\rho_{3}$, with a concomitant reduction of all doublon probabilities $\rho_{2,\Gamma}$ due to Eq.\ (\ref{eq:ProbabilityDensity}).
The sudden increase is followed by a damped oscillatory decay, see Fig.~\ref{fig:DopingDamping}. The corresponding photo-doping $\rho_d$ is defined as the combined change in singlon-triplon probability $\Delta \rho_{1,3}(t) = \rho_{1,3}(t) - \rho_{1,3}(0)$ and holon-quadruplon probability $\Delta \rho_{0,4}(t) = \rho_{0,4}(t) - \rho_{0,4}(0)$, i.e.\ $\rho_d \equiv \Delta \rho_{1,3}(t) + \Delta \rho_{0,4}(t)$. However, since the density of holon-quadruplon high-energy excitations is negligible, $\Delta \rho_{0,4}(t) \approx 0$. Except at very high initial temperature $\rho_d$ effectively measures the singlon-triplon doping $\rho_d \approx \Delta \rho_{1,3}$.
The oscillation frequency $\Omega_{1,3}$ is approximately equal to the Hubbard-$U$, $\Omega_{1,3} \sim U$, and the inverse relaxation time $\tau^{-1}_{1,3}$ is of the order of the bandwidth, $\tau^{-1}_{1,3} \approx W$. Both frequency and damping are only weakly dependent on the Hund's coupling $J$, since the corresponding timescale is long, see the lower panel in Fig.~\ref{fig:DopingDamping}.

The time evolution of the system's energy components -- in the same time window -- is shown in Figs.~\ref{fig:EnergyConversion}a and \ref{fig:EnergyConversion}b. The hopping modulation of the pump only lasts for $t \lesssim 0.5$ and concomitantly the total energy $E_\text{tot}$ of the system displays an oscillatory increase (Fig.~\ref{fig:EnergyConversion}a). After the pulse the total energy $E_\text{tot}$ again becomes a conserved quantity, $\partial_t E_\text{tot}(t) = 0$ for $t \gtrsim 0.5$.
The increase of the photo-doping yields an increase in the local interaction energy $E_\text{int}$. The oscillatory behavior with frequency $\Omega_{1,3} \sim U$ is driven by a shuffling between singlon-triplon interaction energy $E_{1,3}$ and kinetic energy $E_\text{kin}$ at times $t \lesssim 2$, as shown in Fig.~\ref{fig:EnergyConversion}b.

%
The photo-doping (i.e.\ the change in the singlon-triplon probability density $\Delta \rho_{1,3}$) only becomes an approximately conserved quantity after the $2\pi/U$-oscillations have damped out at times $t \gtrsim 2$.
The short-time behavior and photo-doping conservation is similar to the dynamics seen in the single-band Hubbard model. However, at longer times the (paramagnetic) single-band and two-band models exhibit different relaxation dynamics. 

While both the photo-excited single-band and two-band models are (in the large-$U$ limit) trapped in a long lived non-thermal state, since the recombination of particle-hole pairs is exponentially suppressed by the Mott gap \cite{Strohmaier:2010aa, Eckstein:2011aa}, the evolution of the photo-doped population within the Hubbard bands is very different. In the single-band case, the redistribution of spectral weight (and the associated relaxation of the kinetic energy) is very slow \cite{Eckstein:2011aa}. In contrast, as shown in Fig.~\ref{fig:Schematic} and discussed in Sec.~\ref{sec:tdpes}, the photo-doped population can quickly relax within the Hubbard bands due to the additional orbital degrees of freedom and the local spin excitations generated by the Hund's coupling $J$ energy scale.

If we would consider an antiferromagnetically ordered system, then even the single-band Hubbard model would show a similar kinetic energy relaxation \cite{Werner:2012aa}, whose  relaxation rate would be determined by the exchange energy scale $J_\text{ex}$, which is different from the Hund's coupling $J$ that controls the cooling dynamics of the (paramagnetic) two-orbital system.
While we will not consider the interplay of the two scattering processes in the antiferromagnetically ordered two-orbital case in this work, it is clear that the $J_\text{ex}$-based relaxation mechanism will depend strongly on temperature and (within single-site DMFT) will disappear above the Neel temperature, while the Hund's-$J$ based mechanism will persist above this temperature, provided that $J>J_\text{ex}$. Since the flipping of $S=1$ moments requires four hoppings, we generically expect that the Hund's coupling mechanism indeed dominates. 

\subsubsection{Doublon excitations}
\label{sec:doublonExcitations}

The photo-doped two-band Mott-insulator with finite Hund's coupling $|J|>0$ has, due to the lifted doublon degeneracy, internal degrees of freedom in the doublon subsector.
The approximate conservation of the photo-excited singlon-triplon density $\rho_{1,3}(t)$ and holon-quadruplon density $\rho_{0,4}(t)$, i.e.\
\begin{equation}
  \partial_t \rho_{1,3}(t) \approx 0
  \, , \quad
  \partial_t \rho_{0,4}(t) \approx 0
  \label{eq:MottConservation}
\end{equation}
implies that the conservation of total local probability [Eq.\ (\ref{eq:ProbabilityDensity})] constrains the probabilities of the three inequivalent doublon states,
%
%
\begin{equation}
  \partial_t \rho_{2,[\uparrow, \uparrow]} +
  \partial_t \rho_{2,[\uparrow, \downarrow]} +
  \partial_t \rho_{2,[\uparrow\downarrow,0]}
  \approx 0
  \, .
  \label{eq:DoublonProbability}
\end{equation}
%
%
Note that each label is assumed to encompass all possible permutations and directions of spin, i.e. $[\uparrow, \uparrow]$ also includes $[\downarrow, \downarrow]$ and $[\uparrow, \downarrow]$ also includes $[\downarrow, \uparrow]$, etc..

From the energetic point of view the remaining degree of freedom is the conversion between kinetic energy $E_\text{kin}$ and interaction energy in the doublon subsector $E_{\text{int},2}$, i.e.\ $\partial_t (E_\text{kin} + E_{\text{int}, 2}) \approx 0$, where $E_{\text{int},2}$ is given by the weighted sum of the occupation probabilities of the doublon states (with two electrons $N_e=2$) $\rho_{2,\Gamma}$ and their atomic multiplet energies $E_{2,\Gamma}$
%
%
\begin{equation}
  E_\text{int,2} = \sum_{\Gamma \in \{\Gamma : N_e=2\}}
  \rho_{\Gamma} E_{\Gamma}
  \, , \label{eq:DoublonEnergy}
\end{equation}
%
%
c.f.\ Eq.\ (\ref{eq:Eint}).
The kinetic energy relaxation of the pump-generated singlon-triplon excitations by the production of doublon spin excitations, i.e.\ step (II) in the illustration of Fig.\ \ref{fig:Schematic}, can directly be seen in the time-evolution of the system's energy components at times $t \gtrsim 2$ in Fig.~\ref{fig:EnergyConversion}c. The kinetic energy $E_\text{kin}$ shows an exponential decrease, accompanied by a corresponding increase in the interaction energy for the high-energy $S=0$ doublon state with local double occupancy $\Delta E_{2, [\uparrow \downarrow, 0]}$, and a smaller and slower increase in the low-energy $S=0$ doublon state interaction energy $\Delta E_{2,[\uparrow, \downarrow]}$. These two processes correspond to the scattering processes (IIa) and (IIb) in Fig.~\ref{fig:Schematic}, respectively.

\subsubsection{Temperature dependence}
\label{sec:Tsweep}

We find that the speed and magnitude of the response of the different energy components is very sensitive to both the Hund's coupling $J$ and the temperature $T$ of the initial state. Since the temperature dependence is experimentally accessible, we begin by investigating how the relaxation dynamics depends on $T$.

At low temperature $T \ll J$ the initial state is dominated by high spin doublons $[\uparrow, \uparrow]$ and singlon-triplon quantum fluctuations. In this regime the excited doublon states ($[\uparrow,\downarrow]$ and $[\uparrow\downarrow, 0]$) are exponentially suppressed due to the Hund's coupling induced splitting of these states ($J$ and $3J$ respectively).

\begin{figure}
%
%
  \includegraphics[scale=1]
  {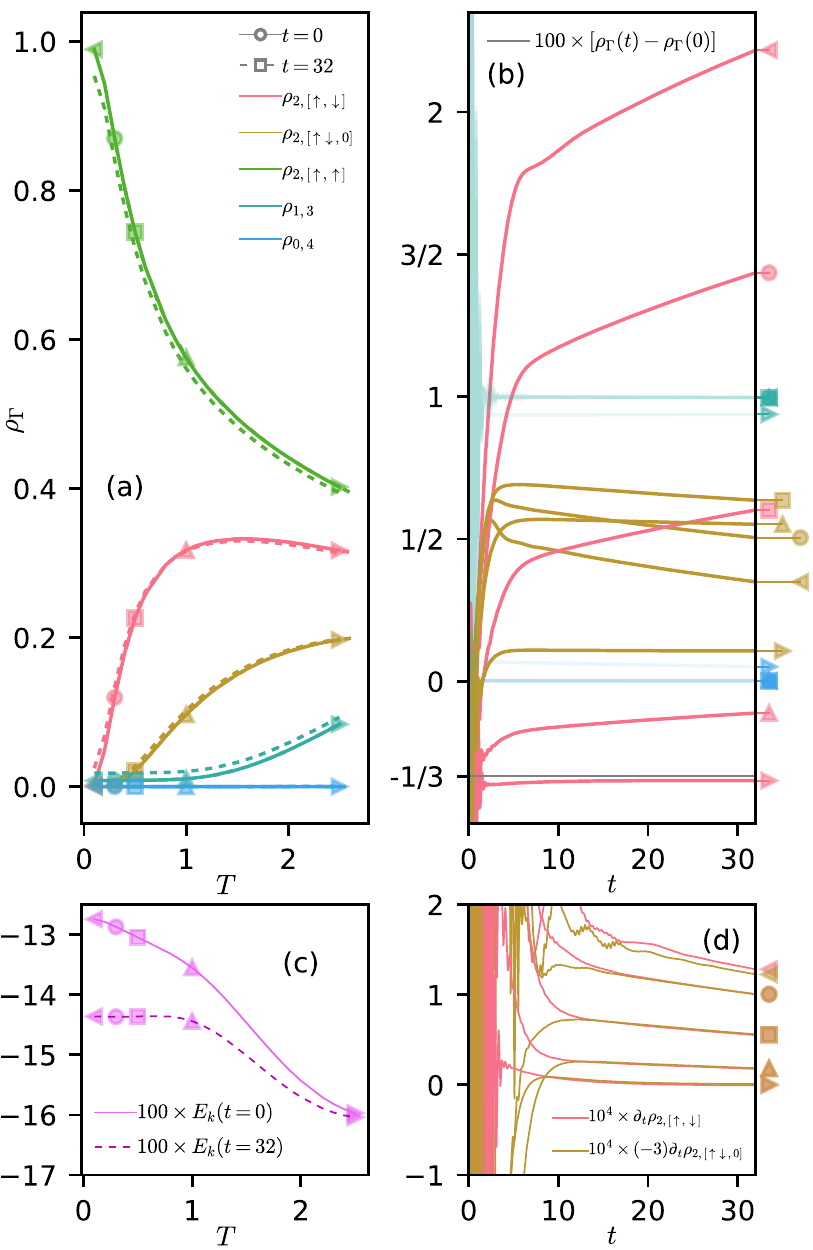} \\[-0mm]
  \caption{\label{fig:Tsweep}(Color online) Panel (a): Local multiplet probabilities $\rho_\Gamma$ at time $t=0$ (solid lines) and at $t=32$ after $1\%$ photo-doping (dashed lines) for a range of temperatures $T=0.1$ to $2.5$ with $W=4$, $U=15$, and $J/U=0.04$, (b) their relative changes $\rho_\Gamma(t) - \rho_\Gamma(0)$ in time, showing the processes IIa (yellow) and IIb (red) of Fig.\ \ref{fig:Schematic}, (c) the kinetic energy $E_k$ at $t=0$ and $t=32$, and (d) scaled time derivatives of the two doublon excitation multiplet densities. The shape of the markers indicates the temperature. In (b) and (d) real-time results for initial temperatures $T=0.1$ (left triangles), $0.3$ (circles), $0.5$ (squares), $1.0$ (up triangles), and $2.5$ (right triangles) are shown.
  }
\end{figure}

To see this we fix $J/U = 0.04$, i.e.\ $U=15$ and $J=0.6$, and sweep temperature from $T=0.1$ to $2.5$, see Fig. \ref{fig:Tsweep}. At $T=0.1$ the high-spin doublon (green circles) dominates the local many body density matrix $\rho_{2,[\uparrow, \uparrow]} \lesssim 1$, while the occupations of the low energy (red circles) and high energy doublons (yellow circles) are exponentially suppressed.
The low energy doublon occupation $\rho_{2,[\uparrow, \downarrow]}$ (red), with an energy splitting of $J=0.6$, rapidly grows when increasing the temperature $T$ from $0.1$ to $1$, see Fig.~\ref{fig:Tsweep}a, while the high energy doublon density $\rho_{2,[\uparrow\downarrow, 0]}$ (yellow) display a slower increase, due to its larger energy splitting $3J = 1.8$.

We find that the $1\%$ photo doping -- i.e.\ the change in the singlon-triplon and holon-quadruplon occupation after the pump, $\rho_p = \Delta \rho_{1,3} + \Delta \rho_{0,4}$ -- does not drastically change the local multiplet occupations. As can be seen from the occupations at time $t=32$ after the pulse (dashed lines in Fig.\ \ref{fig:Tsweep}a), the pump-induced redistribution in the doublon subsector is also of the order of one percent. 

%
%
%

While the absolute changes in the probability distribution among the local multiplet states $\Gamma$ are small, the time dependent deviations from the initial state $\Delta \rho_\Gamma(t) \equiv \rho_\Gamma(t) - \rho_\Gamma(0)$ are non-trivial, see Fig.\ \ref{fig:Tsweep}b. Due to the large Mott gap and the induced conservation laws Eqs.\ (\ref{eq:MottConservation}) and (\ref{eq:DoublonProbability}) the three state dynamics in the doublon sector is captured by the occupation probabilities of the two excited doublons.
The excited doublon densities (red and yellow) show an initial fast exponential rise followed by a slow near linear drift, see  Fig.\ \ref{fig:Tsweep}b. The speed in the initial rise of the high energy doublon (yellow) as compared to the low energy one (red) correlates well with the difference in their multiplet energies ($J$ and $3J$).
Intriguingly, when extrapolating the exponential rise back to $t=0$ (disregarding the $2\pi/U$ oscillations at short times) we find that all doublon densities extrapolate to a $-1/3\%$ absolute reduction as compared to the initial state. In other words, the $1\%$ photo-doping increase of the singlon-triplon and holon-quadruplon states by the high-frequency pulse results in a rapid and even reduction of $1/3\%$ of all three doublon states. This, however, requires all doublon states to be populated in the initial state, i.e. $J \lesssim T$.

%
%
%
\begin{figure}
  \includegraphics[scale=1]
  {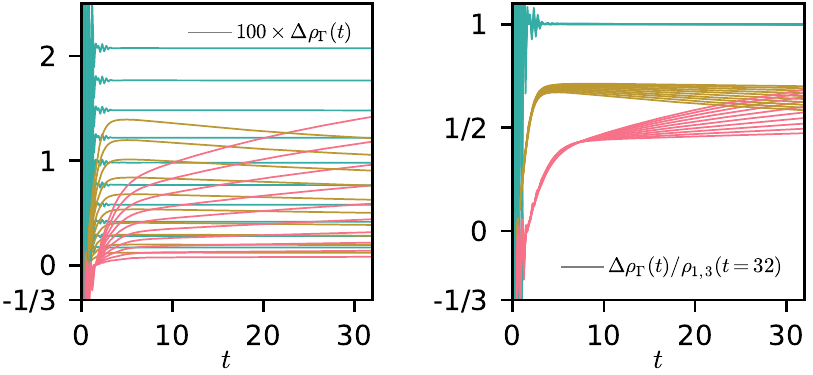} \\[-0mm]
  \caption{\label{fig:Msweep}(Color online) Change in the local multiplet probability relative to the initial ground state, $\Delta \rho_\Gamma(t) \equiv \rho_\Gamma(t) - \rho_\Gamma(0)$, for several different photo-doping densities (left panel). Rescaling with the long-time singlon-triplon photo-doping density $\rho_{1,3}(t=32)$ reveals more clearly the doping dependence of the low spin doublon (red and yellow) conversion rate (right panel). The color coding is defined in Fig.\ \ref{fig:Tsweep}.
  }
\end{figure}
%

The slower dynamics at times $t \gtrsim 5$ additionally conserves the local interaction and kinetic energy, which can be seen in the weighted time derivatives of the doublon occupations, see Fig.\ \ref{fig:Tsweep}d. Every high-energy doublon with energy $3J$ decays into three low energy doublons with energy $J$ by scattering of the high-spin doublon background. This conversion corresponds to the derivative matching: $\partial_t \rho_{2,[\uparrow, \downarrow]} \approx -3 \partial_t \rho_{2,[\uparrow\downarrow, 0]}$ shown in Fig.~\ref{fig:Tsweep}d.
The dominant decay channel for this energy conserving doublon conversion is, however, still mediated by the singlon-triplon photo-doped carriers. Thus the doublon conversion rate is sensitive to the photo-doping density, as shown in Fig.\ \ref{fig:Msweep}.

While the pump generates singlon-triplon pairs with high kinetic energy, the initial relaxation of the total kinetic energy occurs during the initial fast rise for $t \lesssim 5$ (not shown). The largest kinetic energy change ($\approx 10\%$) occurs at low initial temperatures, see Fig.\ \ref{fig:Tsweep}c, and it is substantial considering the $1\%$ fixed photo-doping.
At higher temperature the photo-induced kinetic energy change is reduced since the initial thermal state already populates the excited doublon sector.
In fact, the thermal activation of the low energy doublon results in a \emph{thermal blocking} of its dynamics. This can be seen from the monotonic reduction of the low energy doublon density (red) in Fig.\ \ref{fig:Tsweep}b. At low temperature the dynamics is dominated by the low energy doublon, but as temperature is increased its amplitude is rapidly suppressed (correlating with its thermal activation in the initial state, see Fig.\ \ref{fig:Tsweep}a). For temperatures $T > J = 0.8$ this reduction is so severe that the high-energy doublon (yellow) exhibits the largest response to the photo-doping.

Interestingly, at $T = 2.5$, where the low energy doublon is almost completely thermally blocked, the initial $-1/3\%$ offset still persist. This can be understood as an effect of a quasi instantaneous $1\%$ photo doping change in the singlon-triplon density which by probability conservation [Eq.\ \ref{eq:ProbabilityDensity}] requires the doublon probability densities to be reduced by an equal amount (assuming that the excitation is below the holon-quadruplon threshold).

\subsubsection{Hund's coupling dependence}
\label{sec:Jsweep}

%
%
%
\begin{figure}
  \includegraphics[scale=1]
  {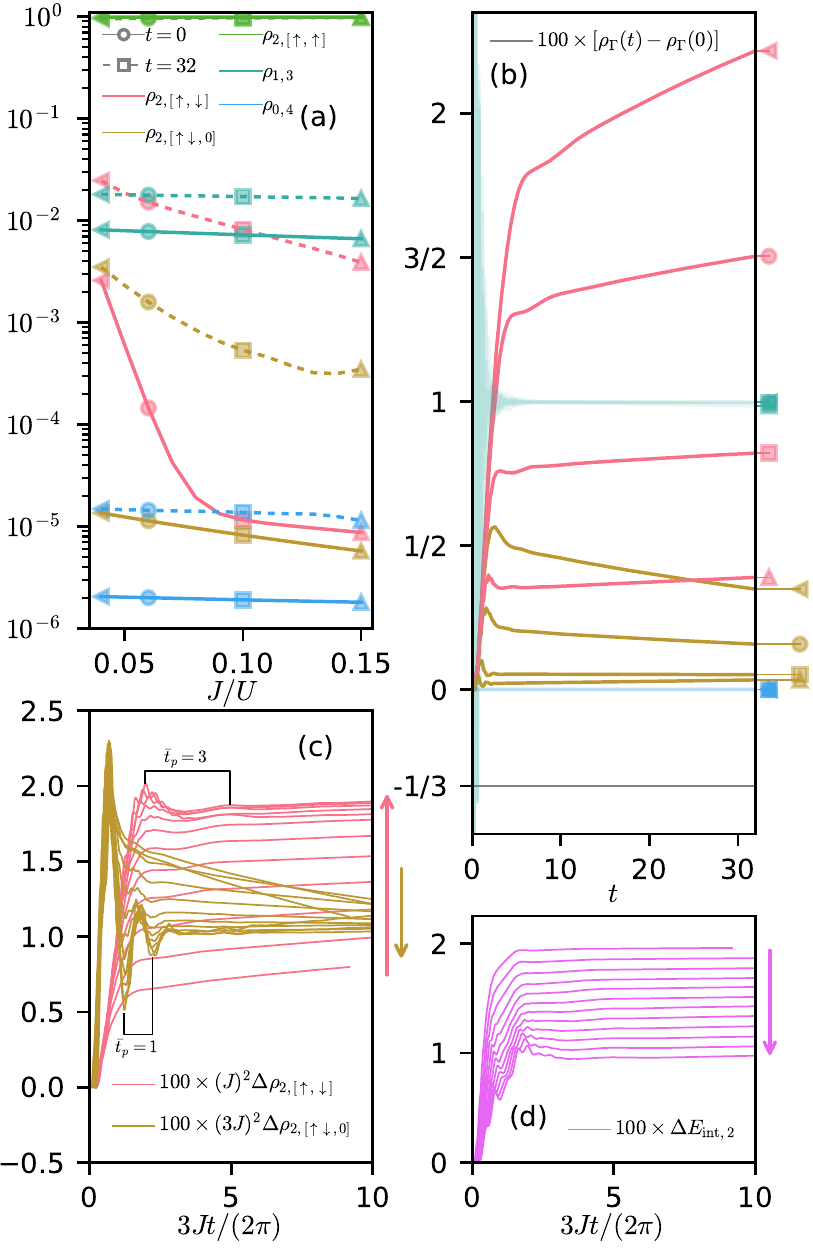} \\[-0mm]
  \caption{\label{fig:Jsweep}(Color online) Panel (a): Local multiplet probabilities $\rho_\Gamma$ at time $t=0$ (solid lines) and at $t=32$ after $1\%$ photo-doping (dashed lines) for a range of Hund's couplings $J/U=0.04$ to $0.15$ with $W=4$, $U=15$ and $\beta=10$, (b) their relative changes $\rho_\Gamma(t) - \rho_\Gamma(0)$ in time, showing the processes IIa (yellow) and IIb (red) of Fig.\ \ref{fig:Schematic} for $J/U=0.04$ (left triangles), $0.06$ (circles), $0.10$ (squares), and $0.15$ (up triangles), (c) rescaled time $2Jt/(2\pi)$ and magnitude $(E_\Gamma)^2 \rho_{2,\Gamma}$ of the two doublon densities, and (d) total change of interaction energy in the doublon subsector $\Delta E_{\text{int},2}$. The color-coded markers indicate the value of the Hund's coupling, and the arrows in (c) and (d) indicate the direction of increasing $J/U$.
  }
\end{figure}
%

We finally investigate the dependence of the photo-carrier relaxation on the Hund's coupling in Fig.\ \ref{fig:Jsweep} at fixed $T=0.1$. At this low initial temperature the doublon excitations are suppressed, while the singlon-triplon fluctuations persist at almost one percent in the initial state, due to hopping driven quantum fluctuations. However, one can still see thermal activation effects of doublons at low $J/U$ since in this limit $T \sim J$. For this reason, the low energy doublon excitation (red circles) displays a marked increase for $J/U \lesssim 0.08$, see Fig.\ \ref{fig:Jsweep}a.

The time-dependence of the probability density changes in Fig.\ \ref{fig:Jsweep}b show that the roles of the doublon states when tuning $J/U$ are reversed, as compared to the increasing temperature. For higher $J/U$ the high-energy doublon state (yellow) is successively more suppressed, and the dynamics is instead dominated by the low energy doublon (red).
The suppression of the high-energy doublon can be understood by kinetic energy arguments. At $J/U \gtrsim 0.09$ the high-energy doublon splitting $3J$ exceeds the bandwidth $W=4$ and high kinetic energy singlons and triplons can no longer excite these high energy doublons. We coin this effect \emph{kinetic freezing}. It  acts in the opposite direction as compared to \emph{thermal blocking}.

There is also an interesting structure in the short time dynamics of the doublon excitations, which directly reveals their atomic-like properties. This manifests itself in the damped oscillations occurring directly after the initial fast rise.
We find that this short time dynamics exhibits a well defined scaling with respect to the Hund's coupling. The time and amplitude rescaling
\begin{equation}
  t \rightarrow \frac{3J}{2\pi} t
  \, , \quad
  \Delta \rho_\Gamma \rightarrow (E_\Gamma)^2 \Delta \rho_\Gamma
\end{equation}
approximately collapses the short time doublon dynamics, see Fig.\ \ref{fig:Jsweep}c. 
This time rescaling reveals that the frequency $\Omega$ of the damped oscillatory response scales with Hund's $J$. The rescaled period $\bar{t}_p = 1$ for the high-energy doublon (yellow) in Fig.\ \ref{fig:Jsweep}c corresponds to a frequency $\Omega = 3J$ while the period $\bar{t} = 3$ for the low energy doublon (red) corresponds to a frequency $\Omega = J$. Thus, the short time response 
contains information on the multiplet splittings. 

The amplitude scaling with the square of the multiplet energy $(E_\Gamma)^2 \propto J^2$ approximately collapses the probability densities for the high-energy doublon (yellow). However, the low-energy doublon dynamics (red) shows clear deviations [Fig.\ \ref{fig:Jsweep}c]. This deviation from the $J^2$ scaling of the low energy doublon occurs in the low $J/U$ regime ($J/U < 0.08$), where it is thermally activated in the initial state, see Fig.\ \ref{fig:Jsweep}a.

After the initial short time dynamics $t \lesssim 10$ the system still shows a conversion between doublon states, while the total interaction and kinetic energy is conserved, see Fig.\ \ref{fig:Jsweep}d. This is a direct analog to the temperature sweep results discussed previously.
Interestingly, we find that the conversion between high and low energy doublon excitations changes direction, when going from low to high $J/U$. While the high energy doublon population indeed gets frozen when $3J > W$, at even higher $J/U$ it starts to slowly grow with time. We speculate that this effect has to do with direct high-energy doublon to singlon-triplon transitions, since at $J/U=0.2$ these two states become degenerate.
In this regime the system becomes unstable with respect to the charge disproportionation \cite{Strand:2014ab} which should have profound effects on the non-equilibrium dynamics. However, since this requires rather high $J/U$ that are not commonly found in real materials we 
do not study this regime here.

While the preceding analysis of the relaxation dynamics has been based on the single-particle spectral function and the local many-body density-matrix, the pseudo-particle method provides a complementary view on the system's dynamics in terms of the pseudo-particle spectral functions. In particular from this analysis it is evident that the dominant contribution to the kinetic energy, and the kinetic energy relaxation, comes from the singlon-triplon states. For details see Appendix~\ref{app:PPRelax}.

\section{Discussion and conclusions}
\label{sec:discussionAndConclusion}

A series of recent theoretical studies has shown that already the single-band Hubbard model exhibits a non-trivial thermalization dynamics after photo-excitation \cite{Aoki:2014kx, Kemper:2016aa, Strohmaier:2010aa}, in particular an exponential dependence of the photo-carrier lifetime with the Mott gap \cite{Eckstein:2011aa, Lenarifmmode-celse-cfiiifmmode-celse-cfi:2013aa, Strohmaier:2010aa}, impact ionization in the small gap regime $U \gtrsim W$ \cite{Werner:2014aa}, and an unconventional nature of the photo-doped metallic state \cite{Eckstein:2013aa}.
Non-local interactions \cite{0295-5075-109-3-37002, Golez:2017aa,  doi:10.1063/1.4935245} and spin correlations \cite{Werner:2012aa, Goleifmmode-zelse-zfi:2015ab} lead to qualitative changes in the relaxation dynamics of the photo-carriers, as does the coupling to phonons \cite{0295-5075-109-3-37002, Dorfner:2015aa, Sayyad:2015aa, Murakami:2015aa, Goleifmmode-zelse-zfi:2012aa}.

The results of the present analysis show that local orbital degrees of freedom and Hund's coupling further enrich the response to photo-doping in the strongly interacting Mott insulator, and lead to qualitatively new physics, which manifests itself in new features of the nonequilibrium spectral function and new relaxation timescales. 
This is central for the understanding of pump-probe experiments, since most correlated materials contain multiple orbitals in the vicinity of the Fermi level (typically a mix of transition metal $d$-states and oxygen/pnictogen $p$-orbitals) and cannot be represented by an effective single-band Hubbard model. 
Here, we focused on the case with strongly correlated orbitals only. The analysis of a photo-doped state with weakly correlated holes and strongly correlated doublons, as realized for example in a $d$-$p$ model of cuprates, is an interesting topic for a separate future study.

To highlight the qualitative differences between the single- and multi-band systems we have studied the dynamics of the paramagnetic half-filled two-band Hubbard model with local density-density interactions after a photo-doping pulse.
By analyzing the dynamics of the local many body states and the single-particle time-dependent photo-emission spectra (TD-PES), we identified a number of qualitatively new features of multi-orbital models.
In particular we observed additional side peaks in the occupied TD-PES spectra, and showed that these result from transitions between local states with one and two electrons less than the nominal filling. This is a general feature of multi-orbital systems with several local occupation states.
We also observed a splitting of the photo-doped carrier population in the upper Hubbard band in the TD-PES spectra, and showed that this is a direct effect of the Hund's coupling. The experimental observation of such a dynamic splitting would be a ``smoking gun'' experiment for non-equilibrium Hund's physics.

By studying the time evolution of the local many-body states we also identified two distinct multi-orbital Hund's coupling effects, which we denoted \emph{thermal blocking} and \emph{kinetic freezing}.
Thermal blocking in the photo-doping dynamics arises when the temperature of the initial state competes with the Hund's splitting of the local states, $T \gtrsim J$, inhibiting the dynamics of thermally occupied local spin excitations.
The kinetic freezing effect on the other hand is the competition between the Hund's coupling $J$ and the bandwidth $W$ of the electronic hopping. In the regime $W \lesssim J$ the internal dynamics of the local spin excitations are frozen out since the available kinetic energy is less than the energy $J$ or $3J$ required to generate such spin excitations.

An interesting venue for further studies is the photo-doping of the anti-ferromagnetic state, and its characteristic changes with temperature. 
In this case, there is an additional energy scale (the antiferromagnetic exchange coupling $J_\text{ex}$), and the relaxation dynamics will depend on the relative values of $T$, $W$, $J$ and $J_\text{ex}$.
It should also be interesting to investigate the quantitative changes in the dynamics of the two-band model when going from density-density interaction case studied here to the rotationally invariant Kanamori interaction (setting $\gamma=1$ in Eq.\ [\ref{eq:HlocHund}]). Since this changes the degeneracy of the doublon states and modifies the multiplet splittings \cite{Antipov:2012aa}, from $J$ \& $3J$ to $2J$ \& $4J$, we expect the onset of \emph{thermal blocking} and \emph{kinetic freezing} to shift.

An interesting material to explore some of the Hund's coupling effects discussed in this paper is the paramagnetic Mott insulator Ca$_2$RuO$_4$. 
A recent ARPES study of this material, and a combined density functional theory plus DMFT investigation \cite{Sutter:2016aa} suggest that spectral features originating from the Hund's coupling are present already in equilibrium. Since the proposed effective local interaction parameters for this system yield $J/U \approx 0.17$ we predict that a time-dependent photo emission study with pump excitation across the Mott gap of $\approx \!\! 2.7$ eV should produce a pronounced splitting in the photo-carrier distribution in the upper Hubbard band, qualitatively similar to what we observe in Fig.\ \ref{fig:tdpes}c-d.

\begin{acknowledgments}
The calculations have been performed on the UniFr cluster. H.U.R.S., D.G., and P.W. were supported by the European Research Council under the European Union's Seventh Framework Programme (FP7/2007-2013) / ERC Grant Agreement No.~278023 and the Swiss National Science Foundation (Grant No.~200021-140648).
\end{acknowledgments}

 \bibliography{/Users/hugstr/Documents/Papers/DMFT_Biblography}

\begin{thebibliography}{74}%
\makeatletter
\providecommand \@ifxundefined [1]{%
 \@ifx{#1\undefined}
}%
\providecommand \@ifnum [1]{%
 \ifnum #1\expandafter \@firstoftwo
 \else \expandafter \@secondoftwo
 \fi
}%
\providecommand \@ifx [1]{%
 \ifx #1\expandafter \@firstoftwo
 \else \expandafter \@secondoftwo
 \fi
}%
\providecommand \natexlab [1]{#1}%
\providecommand \enquote  [1]{``#1''}%
\providecommand \bibnamefont  [1]{#1}%
\providecommand \bibfnamefont [1]{#1}%
\providecommand \citenamefont [1]{#1}%
\providecommand \href@noop [0]{\@secondoftwo}%
\providecommand \href [0]{\begingroup \@sanitize@url \@href}%
\providecommand \@href[1]{\@@startlink{#1}\@@href}%
\providecommand \@@href[1]{\endgroup#1\@@endlink}%
\providecommand \@sanitize@url [0]{\catcode `\\12\catcode `\$12\catcode
  `\&12\catcode `\#12\catcode `\^12\catcode `\_12\catcode `\%12\relax}%
\providecommand \@@startlink[1]{}%
\providecommand \@@endlink[0]{}%
\providecommand \url  [0]{\begingroup\@sanitize@url \@url }%
\providecommand \@url [1]{\endgroup\@href {#1}{\urlprefix }}%
\providecommand \urlprefix  [0]{URL }%
\providecommand \Eprint [0]{\href }%
\providecommand \doibase [0]{http://dx.doi.org/}%
\providecommand \selectlanguage [0]{\@gobble}%
\providecommand \bibinfo  [0]{\@secondoftwo}%
\providecommand \bibfield  [0]{\@secondoftwo}%
\providecommand \translation [1]{[#1]}%
\providecommand \BibitemOpen [0]{}%
\providecommand \bibitemStop [0]{}%
\providecommand \bibitemNoStop [0]{.\EOS\space}%
\providecommand \EOS [0]{\spacefactor3000\relax}%
\providecommand \BibitemShut  [1]{\csname bibitem#1\endcsname}%
\let\auto@bib@innerbib\@empty
\bibitem [{\citenamefont {Iwai}\ \emph {et~al.}(2003)\citenamefont {Iwai},
  \citenamefont {Ono}, \citenamefont {Maeda}, \citenamefont {Matsuzaki},
  \citenamefont {Kishida}, \citenamefont {Okamoto},\ and\ \citenamefont
  {Tokura}}]{Iwai:2003aa}%
  \BibitemOpen
  \bibfield  {author} {\bibinfo {author} {\bibfnamefont {S.}~\bibnamefont
  {Iwai}}, \bibinfo {author} {\bibfnamefont {M.}~\bibnamefont {Ono}}, \bibinfo
  {author} {\bibfnamefont {A.}~\bibnamefont {Maeda}}, \bibinfo {author}
  {\bibfnamefont {H.}~\bibnamefont {Matsuzaki}}, \bibinfo {author}
  {\bibfnamefont {H.}~\bibnamefont {Kishida}}, \bibinfo {author} {\bibfnamefont
  {H.}~\bibnamefont {Okamoto}}, \ and\ \bibinfo {author} {\bibfnamefont
  {Y.}~\bibnamefont {Tokura}},\ }\href {\doibase 10.1103/PhysRevLett.91.057401}
  {\bibfield  {journal} {\bibinfo  {journal} {Phys. Rev. Lett.}\ }\textbf
  {\bibinfo {volume} {91}},\ \bibinfo {pages} {057401} (\bibinfo {year}
  {2003})}\BibitemShut {NoStop}%
\bibitem [{\citenamefont {Okamoto}\ \emph {et~al.}(2010)\citenamefont
  {Okamoto}, \citenamefont {Miyagoe}, \citenamefont {Kobayashi}, \citenamefont
  {Uemura}, \citenamefont {Nishioka}, \citenamefont {Matsuzaki}, \citenamefont
  {Sawa},\ and\ \citenamefont {Tokura}}]{Okamoto:2010aa}%
  \BibitemOpen
  \bibfield  {author} {\bibinfo {author} {\bibfnamefont {H.}~\bibnamefont
  {Okamoto}}, \bibinfo {author} {\bibfnamefont {T.}~\bibnamefont {Miyagoe}},
  \bibinfo {author} {\bibfnamefont {K.}~\bibnamefont {Kobayashi}}, \bibinfo
  {author} {\bibfnamefont {H.}~\bibnamefont {Uemura}}, \bibinfo {author}
  {\bibfnamefont {H.}~\bibnamefont {Nishioka}}, \bibinfo {author}
  {\bibfnamefont {H.}~\bibnamefont {Matsuzaki}}, \bibinfo {author}
  {\bibfnamefont {A.}~\bibnamefont {Sawa}}, \ and\ \bibinfo {author}
  {\bibfnamefont {Y.}~\bibnamefont {Tokura}},\ }\href {\doibase
  10.1103/PhysRevB.82.060513} {\bibfield  {journal} {\bibinfo  {journal} {Phys.
  Rev. B}\ }\textbf {\bibinfo {volume} {82}},\ \bibinfo {pages} {060513}
  (\bibinfo {year} {2010})}\BibitemShut {NoStop}%
\bibitem [{\citenamefont {Perfetti}\ \emph {et~al.}(2006)\citenamefont
  {Perfetti}, \citenamefont {Loukakos}, \citenamefont {Lisowski}, \citenamefont
  {Bovensiepen}, \citenamefont {Berger}, \citenamefont {Biermann},
  \citenamefont {Cornaglia}, \citenamefont {Georges},\ and\ \citenamefont
  {Wolf}}]{Perfetti:2006aa}%
  \BibitemOpen
  \bibfield  {author} {\bibinfo {author} {\bibfnamefont {L.}~\bibnamefont
  {Perfetti}}, \bibinfo {author} {\bibfnamefont {P.~A.}\ \bibnamefont
  {Loukakos}}, \bibinfo {author} {\bibfnamefont {M.}~\bibnamefont {Lisowski}},
  \bibinfo {author} {\bibfnamefont {U.}~\bibnamefont {Bovensiepen}}, \bibinfo
  {author} {\bibfnamefont {H.}~\bibnamefont {Berger}}, \bibinfo {author}
  {\bibfnamefont {S.}~\bibnamefont {Biermann}}, \bibinfo {author}
  {\bibfnamefont {P.~S.}\ \bibnamefont {Cornaglia}}, \bibinfo {author}
  {\bibfnamefont {A.}~\bibnamefont {Georges}}, \ and\ \bibinfo {author}
  {\bibfnamefont {M.}~\bibnamefont {Wolf}},\ }\href {\doibase
  10.1103/PhysRevLett.97.067402} {\bibfield  {journal} {\bibinfo  {journal}
  {Phys. Rev. Lett.}\ }\textbf {\bibinfo {volume} {97}},\ \bibinfo {pages}
  {067402} (\bibinfo {year} {2006})}\BibitemShut {NoStop}%
\bibitem [{\citenamefont {Perfetti}\ \emph {et~al.}(2008)\citenamefont
  {Perfetti}, \citenamefont {Loukakos}, \citenamefont {Lisowski}, \citenamefont
  {Bovensiepen}, \citenamefont {Wolf}, \citenamefont {Berger}, \citenamefont
  {Biermann},\ and\ \citenamefont {Georges}}]{Perfetti:2008aa}%
  \BibitemOpen
  \bibfield  {author} {\bibinfo {author} {\bibfnamefont {L.}~\bibnamefont
  {Perfetti}}, \bibinfo {author} {\bibfnamefont {P.~A.}\ \bibnamefont
  {Loukakos}}, \bibinfo {author} {\bibfnamefont {M.}~\bibnamefont {Lisowski}},
  \bibinfo {author} {\bibfnamefont {U.}~\bibnamefont {Bovensiepen}}, \bibinfo
  {author} {\bibfnamefont {M.}~\bibnamefont {Wolf}}, \bibinfo {author}
  {\bibfnamefont {H.}~\bibnamefont {Berger}}, \bibinfo {author} {\bibfnamefont
  {S.}~\bibnamefont {Biermann}}, \ and\ \bibinfo {author} {\bibfnamefont
  {A.}~\bibnamefont {Georges}},\ }\href
  {http://stacks.iop.org/1367-2630/10/i=5/a=053019} {\bibfield  {journal}
  {\bibinfo  {journal} {New Journal of Physics}\ }\textbf {\bibinfo {volume}
  {10}},\ \bibinfo {pages} {053019} (\bibinfo {year} {2008})}\BibitemShut
  {NoStop}%
\bibitem [{\citenamefont {{Ligges}}\ \emph {et~al.}(2017)\citenamefont
  {{Ligges}}, \citenamefont {{Avigo}}, \citenamefont {{Gole{\v z}}},
  \citenamefont {{Strand}}, \citenamefont {{Stojchevska}}, \citenamefont
  {{Kall{\"a}ne}}, \citenamefont {{Rossnagel}}, \citenamefont {{Eckstein}},
  \citenamefont {{Werner}},\ and\ \citenamefont
  {{Bovensiepen}}}]{Ligges:2017aa}%
  \BibitemOpen
  \bibfield  {author} {\bibinfo {author} {\bibfnamefont {M.}~\bibnamefont
  {{Ligges}}}, \bibinfo {author} {\bibfnamefont {I.}~\bibnamefont {{Avigo}}},
  \bibinfo {author} {\bibfnamefont {D.}~\bibnamefont {{Gole{\v z}}}}, \bibinfo
  {author} {\bibfnamefont {H.}~\bibnamefont {{Strand}}}, \bibinfo {author}
  {\bibfnamefont {L.}~\bibnamefont {{Stojchevska}}}, \bibinfo {author}
  {\bibfnamefont {M.}~\bibnamefont {{Kall{\"a}ne}}}, \bibinfo {author}
  {\bibfnamefont {K.}~\bibnamefont {{Rossnagel}}}, \bibinfo {author}
  {\bibfnamefont {M.}~\bibnamefont {{Eckstein}}}, \bibinfo {author}
  {\bibfnamefont {P.}~\bibnamefont {{Werner}}}, \ and\ \bibinfo {author}
  {\bibfnamefont {U.}~\bibnamefont {{Bovensiepen}}},\ }\href@noop {} {\bibfield
   {journal} {\bibinfo  {journal} {ArXiv e-prints}\ } (\bibinfo {year}
  {2017})},\ \Eprint {http://arxiv.org/abs/1702.05300} {arXiv:1702.05300
  [cond-mat.str-el]} \BibitemShut {NoStop}%
\bibitem [{\citenamefont {Johnson}\ \emph {et~al.}(2012)\citenamefont
  {Johnson}, \citenamefont {de~Souza}, \citenamefont {Staub}, \citenamefont
  {Beaud}, \citenamefont {M\"ohr-Vorobeva}, \citenamefont {Ingold},
  \citenamefont {Caviezel}, \citenamefont {Scagnoli}, \citenamefont
  {Schlotter}, \citenamefont {Turner}, \citenamefont {Krupin}, \citenamefont
  {Lee}, \citenamefont {Chuang}, \citenamefont {Patthey}, \citenamefont
  {Moore}, \citenamefont {Lu}, \citenamefont {Yi}, \citenamefont {Kirchmann},
  \citenamefont {Trigo}, \citenamefont {Denes}, \citenamefont {Doering},
  \citenamefont {Hussain}, \citenamefont {Shen}, \citenamefont {Prabhakaran},\
  and\ \citenamefont {Boothroyd}}]{Johnson:2012aa}%
  \BibitemOpen
  \bibfield  {author} {\bibinfo {author} {\bibfnamefont {S.~L.}\ \bibnamefont
  {Johnson}}, \bibinfo {author} {\bibfnamefont {R.~A.}\ \bibnamefont
  {de~Souza}}, \bibinfo {author} {\bibfnamefont {U.}~\bibnamefont {Staub}},
  \bibinfo {author} {\bibfnamefont {P.}~\bibnamefont {Beaud}}, \bibinfo
  {author} {\bibfnamefont {E.}~\bibnamefont {M\"ohr-Vorobeva}}, \bibinfo
  {author} {\bibfnamefont {G.}~\bibnamefont {Ingold}}, \bibinfo {author}
  {\bibfnamefont {A.}~\bibnamefont {Caviezel}}, \bibinfo {author}
  {\bibfnamefont {V.}~\bibnamefont {Scagnoli}}, \bibinfo {author}
  {\bibfnamefont {W.~F.}\ \bibnamefont {Schlotter}}, \bibinfo {author}
  {\bibfnamefont {J.~J.}\ \bibnamefont {Turner}}, \bibinfo {author}
  {\bibfnamefont {O.}~\bibnamefont {Krupin}}, \bibinfo {author} {\bibfnamefont
  {W.-S.}\ \bibnamefont {Lee}}, \bibinfo {author} {\bibfnamefont {Y.-D.}\
  \bibnamefont {Chuang}}, \bibinfo {author} {\bibfnamefont {L.}~\bibnamefont
  {Patthey}}, \bibinfo {author} {\bibfnamefont {R.~G.}\ \bibnamefont {Moore}},
  \bibinfo {author} {\bibfnamefont {D.}~\bibnamefont {Lu}}, \bibinfo {author}
  {\bibfnamefont {M.}~\bibnamefont {Yi}}, \bibinfo {author} {\bibfnamefont
  {P.~S.}\ \bibnamefont {Kirchmann}}, \bibinfo {author} {\bibfnamefont
  {M.}~\bibnamefont {Trigo}}, \bibinfo {author} {\bibfnamefont
  {P.}~\bibnamefont {Denes}}, \bibinfo {author} {\bibfnamefont
  {D.}~\bibnamefont {Doering}}, \bibinfo {author} {\bibfnamefont
  {Z.}~\bibnamefont {Hussain}}, \bibinfo {author} {\bibfnamefont {Z.-X.}\
  \bibnamefont {Shen}}, \bibinfo {author} {\bibfnamefont {D.}~\bibnamefont
  {Prabhakaran}}, \ and\ \bibinfo {author} {\bibfnamefont {A.~T.}\ \bibnamefont
  {Boothroyd}},\ }\href {\doibase 10.1103/PhysRevLett.108.037203} {\bibfield
  {journal} {\bibinfo  {journal} {Phys. Rev. Lett.}\ }\textbf {\bibinfo
  {volume} {108}},\ \bibinfo {pages} {037203} (\bibinfo {year}
  {2012})}\BibitemShut {NoStop}%
\bibitem [{\citenamefont {Johnson}\ \emph {et~al.}(2015)\citenamefont
  {Johnson}, \citenamefont {Kubacka}, \citenamefont {Hoffmann}, \citenamefont
  {Vicario}, \citenamefont {de~Jong}, \citenamefont {Beaud}, \citenamefont
  {Gr\"ubel}, \citenamefont {Huang}, \citenamefont {Huber}, \citenamefont
  {Windsor}, \citenamefont {Bothschafter}, \citenamefont {Rettig},
  \citenamefont {Ramakrishnan}, \citenamefont {Alberca}, \citenamefont
  {Patthey}, \citenamefont {Chuang}, \citenamefont {Turner}, \citenamefont
  {Dakovski}, \citenamefont {Lee}, \citenamefont {Minitti}, \citenamefont
  {Schlotter}, \citenamefont {Moore}, \citenamefont {Hauri}, \citenamefont
  {Koohpayeh}, \citenamefont {Scagnoli}, \citenamefont {Ingold}, \citenamefont
  {Johnson},\ and\ \citenamefont {Staub}}]{Johnson:2015aa}%
  \BibitemOpen
  \bibfield  {author} {\bibinfo {author} {\bibfnamefont {J.~A.}\ \bibnamefont
  {Johnson}}, \bibinfo {author} {\bibfnamefont {T.}~\bibnamefont {Kubacka}},
  \bibinfo {author} {\bibfnamefont {M.~C.}\ \bibnamefont {Hoffmann}}, \bibinfo
  {author} {\bibfnamefont {C.}~\bibnamefont {Vicario}}, \bibinfo {author}
  {\bibfnamefont {S.}~\bibnamefont {de~Jong}}, \bibinfo {author} {\bibfnamefont
  {P.}~\bibnamefont {Beaud}}, \bibinfo {author} {\bibfnamefont
  {S.}~\bibnamefont {Gr\"ubel}}, \bibinfo {author} {\bibfnamefont {S.-W.}\
  \bibnamefont {Huang}}, \bibinfo {author} {\bibfnamefont {L.}~\bibnamefont
  {Huber}}, \bibinfo {author} {\bibfnamefont {Y.~W.}\ \bibnamefont {Windsor}},
  \bibinfo {author} {\bibfnamefont {E.~M.}\ \bibnamefont {Bothschafter}},
  \bibinfo {author} {\bibfnamefont {L.}~\bibnamefont {Rettig}}, \bibinfo
  {author} {\bibfnamefont {M.}~\bibnamefont {Ramakrishnan}}, \bibinfo {author}
  {\bibfnamefont {A.}~\bibnamefont {Alberca}}, \bibinfo {author} {\bibfnamefont
  {L.}~\bibnamefont {Patthey}}, \bibinfo {author} {\bibfnamefont {Y.-D.}\
  \bibnamefont {Chuang}}, \bibinfo {author} {\bibfnamefont {J.~J.}\
  \bibnamefont {Turner}}, \bibinfo {author} {\bibfnamefont {G.~L.}\
  \bibnamefont {Dakovski}}, \bibinfo {author} {\bibfnamefont {W.-S.}\
  \bibnamefont {Lee}}, \bibinfo {author} {\bibfnamefont {M.~P.}\ \bibnamefont
  {Minitti}}, \bibinfo {author} {\bibfnamefont {W.}~\bibnamefont {Schlotter}},
  \bibinfo {author} {\bibfnamefont {R.~G.}\ \bibnamefont {Moore}}, \bibinfo
  {author} {\bibfnamefont {C.~P.}\ \bibnamefont {Hauri}}, \bibinfo {author}
  {\bibfnamefont {S.~M.}\ \bibnamefont {Koohpayeh}}, \bibinfo {author}
  {\bibfnamefont {V.}~\bibnamefont {Scagnoli}}, \bibinfo {author}
  {\bibfnamefont {G.}~\bibnamefont {Ingold}}, \bibinfo {author} {\bibfnamefont
  {S.~L.}\ \bibnamefont {Johnson}}, \ and\ \bibinfo {author} {\bibfnamefont
  {U.}~\bibnamefont {Staub}},\ }\href {\doibase 10.1103/PhysRevB.92.184429}
  {\bibfield  {journal} {\bibinfo  {journal} {Phys. Rev. B}\ }\textbf {\bibinfo
  {volume} {92}},\ \bibinfo {pages} {184429} (\bibinfo {year}
  {2015})}\BibitemShut {NoStop}%
\bibitem [{\citenamefont {Allen}(1987)}]{Allen:1987aa}%
  \BibitemOpen
  \bibfield  {author} {\bibinfo {author} {\bibfnamefont {P.~B.}\ \bibnamefont
  {Allen}},\ }\href {\doibase 10.1103/PhysRevLett.59.1460} {\bibfield
  {journal} {\bibinfo  {journal} {Phys. Rev. Lett.}\ }\textbf {\bibinfo
  {volume} {59}},\ \bibinfo {pages} {1460} (\bibinfo {year}
  {1987})}\BibitemShut {NoStop}%
\bibitem [{\citenamefont {Brorson}\ \emph {et~al.}(1990)\citenamefont
  {Brorson}, \citenamefont {Kazeroonian}, \citenamefont {Moodera},
  \citenamefont {Face}, \citenamefont {Cheng}, \citenamefont {Ippen},
  \citenamefont {Dresselhaus},\ and\ \citenamefont
  {Dresselhaus}}]{Brorson:1990aa}%
  \BibitemOpen
  \bibfield  {author} {\bibinfo {author} {\bibfnamefont {S.~D.}\ \bibnamefont
  {Brorson}}, \bibinfo {author} {\bibfnamefont {A.}~\bibnamefont
  {Kazeroonian}}, \bibinfo {author} {\bibfnamefont {J.~S.}\ \bibnamefont
  {Moodera}}, \bibinfo {author} {\bibfnamefont {D.~W.}\ \bibnamefont {Face}},
  \bibinfo {author} {\bibfnamefont {T.~K.}\ \bibnamefont {Cheng}}, \bibinfo
  {author} {\bibfnamefont {E.~P.}\ \bibnamefont {Ippen}}, \bibinfo {author}
  {\bibfnamefont {M.~S.}\ \bibnamefont {Dresselhaus}}, \ and\ \bibinfo {author}
  {\bibfnamefont {G.}~\bibnamefont {Dresselhaus}},\ }\href {\doibase
  10.1103/PhysRevLett.64.2172} {\bibfield  {journal} {\bibinfo  {journal}
  {Phys. Rev. Lett.}\ }\textbf {\bibinfo {volume} {64}},\ \bibinfo {pages}
  {2172} (\bibinfo {year} {1990})}\BibitemShut {NoStop}%
\bibitem [{\citenamefont {Gadermaier}\ \emph {et~al.}(2010)\citenamefont
  {Gadermaier}, \citenamefont {Alexandrov}, \citenamefont {Kabanov},
  \citenamefont {Kusar}, \citenamefont {Mertelj}, \citenamefont {Yao},
  \citenamefont {Manzoni}, \citenamefont {Brida}, \citenamefont {Cerullo},\
  and\ \citenamefont {Mihailovic}}]{Gadermaier:2010aa}%
  \BibitemOpen
  \bibfield  {author} {\bibinfo {author} {\bibfnamefont {C.}~\bibnamefont
  {Gadermaier}}, \bibinfo {author} {\bibfnamefont {A.~S.}\ \bibnamefont
  {Alexandrov}}, \bibinfo {author} {\bibfnamefont {V.~V.}\ \bibnamefont
  {Kabanov}}, \bibinfo {author} {\bibfnamefont {P.}~\bibnamefont {Kusar}},
  \bibinfo {author} {\bibfnamefont {T.}~\bibnamefont {Mertelj}}, \bibinfo
  {author} {\bibfnamefont {X.}~\bibnamefont {Yao}}, \bibinfo {author}
  {\bibfnamefont {C.}~\bibnamefont {Manzoni}}, \bibinfo {author} {\bibfnamefont
  {D.}~\bibnamefont {Brida}}, \bibinfo {author} {\bibfnamefont
  {G.}~\bibnamefont {Cerullo}}, \ and\ \bibinfo {author} {\bibfnamefont
  {D.}~\bibnamefont {Mihailovic}},\ }\href {\doibase
  10.1103/PhysRevLett.105.257001} {\bibfield  {journal} {\bibinfo  {journal}
  {Phys. Rev. Lett.}\ }\textbf {\bibinfo {volume} {105}},\ \bibinfo {pages}
  {257001} (\bibinfo {year} {2010})}\BibitemShut {NoStop}%
\bibitem [{\citenamefont {van Heumen}\ \emph {et~al.}(2009)\citenamefont {van
  Heumen}, \citenamefont {Muhlethaler}, \citenamefont {Kuzmenko}, \citenamefont
  {Eisaki}, \citenamefont {Meevasana}, \citenamefont {Greven},\ and\
  \citenamefont {van~der Marel}}]{Heumen:2009aa}%
  \BibitemOpen
  \bibfield  {author} {\bibinfo {author} {\bibfnamefont {E.}~\bibnamefont {van
  Heumen}}, \bibinfo {author} {\bibfnamefont {E.}~\bibnamefont {Muhlethaler}},
  \bibinfo {author} {\bibfnamefont {A.~B.}\ \bibnamefont {Kuzmenko}}, \bibinfo
  {author} {\bibfnamefont {H.}~\bibnamefont {Eisaki}}, \bibinfo {author}
  {\bibfnamefont {W.}~\bibnamefont {Meevasana}}, \bibinfo {author}
  {\bibfnamefont {M.}~\bibnamefont {Greven}}, \ and\ \bibinfo {author}
  {\bibfnamefont {D.}~\bibnamefont {van~der Marel}},\ }\href {\doibase
  10.1103/PhysRevB.79.184512} {\bibfield  {journal} {\bibinfo  {journal} {Phys.
  Rev. B}\ }\textbf {\bibinfo {volume} {79}},\ \bibinfo {pages} {184512}
  (\bibinfo {year} {2009})}\BibitemShut {NoStop}%
\bibitem [{\citenamefont {Rameau}\ \emph {et~al.}(2016)\citenamefont {Rameau},
  \citenamefont {Freutel}, \citenamefont {Kemper}, \citenamefont {Sentef},
  \citenamefont {Freericks}, \citenamefont {Avigo}, \citenamefont {Ligges},
  \citenamefont {Rettig}, \citenamefont {Yoshida}, \citenamefont {Eisaki},
  \citenamefont {Schneeloch}, \citenamefont {Zhong}, \citenamefont {Xu},
  \citenamefont {Gu}, \citenamefont {Johnson},\ and\ \citenamefont
  {Bovensiepen}}]{Rameau:2016aa}%
  \BibitemOpen
  \bibfield  {author} {\bibinfo {author} {\bibfnamefont {J.~D.}\ \bibnamefont
  {Rameau}}, \bibinfo {author} {\bibfnamefont {S.}~\bibnamefont {Freutel}},
  \bibinfo {author} {\bibfnamefont {A.~F.}\ \bibnamefont {Kemper}}, \bibinfo
  {author} {\bibfnamefont {M.~A.}\ \bibnamefont {Sentef}}, \bibinfo {author}
  {\bibfnamefont {J.~K.}\ \bibnamefont {Freericks}}, \bibinfo {author}
  {\bibfnamefont {I.}~\bibnamefont {Avigo}}, \bibinfo {author} {\bibfnamefont
  {M.}~\bibnamefont {Ligges}}, \bibinfo {author} {\bibfnamefont
  {L.}~\bibnamefont {Rettig}}, \bibinfo {author} {\bibfnamefont
  {Y.}~\bibnamefont {Yoshida}}, \bibinfo {author} {\bibfnamefont
  {H.}~\bibnamefont {Eisaki}}, \bibinfo {author} {\bibfnamefont
  {J.}~\bibnamefont {Schneeloch}}, \bibinfo {author} {\bibfnamefont {R.~D.}\
  \bibnamefont {Zhong}}, \bibinfo {author} {\bibfnamefont {Z.~J.}\ \bibnamefont
  {Xu}}, \bibinfo {author} {\bibfnamefont {G.~D.}\ \bibnamefont {Gu}}, \bibinfo
  {author} {\bibfnamefont {P.~D.}\ \bibnamefont {Johnson}}, \ and\ \bibinfo
  {author} {\bibfnamefont {U.}~\bibnamefont {Bovensiepen}},\ }\href
  {http://dx.doi.org/10.1038/ncomms13761} {\bibfield  {journal} {\bibinfo
  {journal} {Nature Communications}\ }\textbf {\bibinfo {volume} {7}},\
  \bibinfo {pages} {13761 EP } (\bibinfo {year} {2016})}\BibitemShut {NoStop}%
\bibitem [{\citenamefont {Dal~Conte}\ \emph {et~al.}(2012)\citenamefont
  {Dal~Conte}, \citenamefont {Giannetti}, \citenamefont {Coslovich},
  \citenamefont {Cilento}, \citenamefont {Bossini}, \citenamefont {Abebaw},
  \citenamefont {Banfi}, \citenamefont {Ferrini}, \citenamefont {Eisaki},
  \citenamefont {Greven}, \citenamefont {Damascelli}, \citenamefont {van~der
  Marel},\ and\ \citenamefont {Parmigiani}}]{Dal-Conte:2012aa}%
  \BibitemOpen
  \bibfield  {author} {\bibinfo {author} {\bibfnamefont {S.}~\bibnamefont
  {Dal~Conte}}, \bibinfo {author} {\bibfnamefont {C.}~\bibnamefont
  {Giannetti}}, \bibinfo {author} {\bibfnamefont {G.}~\bibnamefont
  {Coslovich}}, \bibinfo {author} {\bibfnamefont {F.}~\bibnamefont {Cilento}},
  \bibinfo {author} {\bibfnamefont {D.}~\bibnamefont {Bossini}}, \bibinfo
  {author} {\bibfnamefont {T.}~\bibnamefont {Abebaw}}, \bibinfo {author}
  {\bibfnamefont {F.}~\bibnamefont {Banfi}}, \bibinfo {author} {\bibfnamefont
  {G.}~\bibnamefont {Ferrini}}, \bibinfo {author} {\bibfnamefont
  {H.}~\bibnamefont {Eisaki}}, \bibinfo {author} {\bibfnamefont
  {M.}~\bibnamefont {Greven}}, \bibinfo {author} {\bibfnamefont
  {A.}~\bibnamefont {Damascelli}}, \bibinfo {author} {\bibfnamefont
  {D.}~\bibnamefont {van~der Marel}}, \ and\ \bibinfo {author} {\bibfnamefont
  {F.}~\bibnamefont {Parmigiani}},\ }\href
  {http://science.sciencemag.org/content/335/6076/1600.abstract} {\bibfield
  {journal} {\bibinfo  {journal} {Science}\ }\textbf {\bibinfo {volume}
  {335}},\ \bibinfo {pages} {1600} (\bibinfo {year} {2012})}\BibitemShut
  {NoStop}%
\bibitem [{\citenamefont {Dal~Conte}\ \emph {et~al.}(2015)\citenamefont
  {Dal~Conte}, \citenamefont {Vidmar}, \citenamefont {Golez}, \citenamefont
  {Mierzejewski}, \citenamefont {Soavi}, \citenamefont {Peli}, \citenamefont
  {Banfi}, \citenamefont {Ferrini}, \citenamefont {Comin}, \citenamefont
  {Ludbrook}, \citenamefont {Chauviere}, \citenamefont {Zhigadlo},
  \citenamefont {Eisaki}, \citenamefont {Greven}, \citenamefont {Lupi},
  \citenamefont {Damascelli}, \citenamefont {Brida}, \citenamefont {Capone},
  \citenamefont {Bonca}, \citenamefont {Cerullo},\ and\ \citenamefont
  {Giannetti}}]{Dal-Conte:2015aa}%
  \BibitemOpen
  \bibfield  {author} {\bibinfo {author} {\bibfnamefont {S.}~\bibnamefont
  {Dal~Conte}}, \bibinfo {author} {\bibfnamefont {L.}~\bibnamefont {Vidmar}},
  \bibinfo {author} {\bibfnamefont {D.}~\bibnamefont {Golez}}, \bibinfo
  {author} {\bibfnamefont {M.}~\bibnamefont {Mierzejewski}}, \bibinfo {author}
  {\bibfnamefont {G.}~\bibnamefont {Soavi}}, \bibinfo {author} {\bibfnamefont
  {S.}~\bibnamefont {Peli}}, \bibinfo {author} {\bibfnamefont {F.}~\bibnamefont
  {Banfi}}, \bibinfo {author} {\bibfnamefont {G.}~\bibnamefont {Ferrini}},
  \bibinfo {author} {\bibfnamefont {R.}~\bibnamefont {Comin}}, \bibinfo
  {author} {\bibfnamefont {B.~M.}\ \bibnamefont {Ludbrook}}, \bibinfo {author}
  {\bibfnamefont {L.}~\bibnamefont {Chauviere}}, \bibinfo {author}
  {\bibfnamefont {N.~D.}\ \bibnamefont {Zhigadlo}}, \bibinfo {author}
  {\bibfnamefont {H.}~\bibnamefont {Eisaki}}, \bibinfo {author} {\bibfnamefont
  {M.}~\bibnamefont {Greven}}, \bibinfo {author} {\bibfnamefont
  {S.}~\bibnamefont {Lupi}}, \bibinfo {author} {\bibfnamefont {A.}~\bibnamefont
  {Damascelli}}, \bibinfo {author} {\bibfnamefont {D.}~\bibnamefont {Brida}},
  \bibinfo {author} {\bibfnamefont {M.}~\bibnamefont {Capone}}, \bibinfo
  {author} {\bibfnamefont {J.}~\bibnamefont {Bonca}}, \bibinfo {author}
  {\bibfnamefont {G.}~\bibnamefont {Cerullo}}, \ and\ \bibinfo {author}
  {\bibfnamefont {C.}~\bibnamefont {Giannetti}},\ }\href
  {http://dx.doi.org/10.1038/nphys3265} {\bibfield  {journal} {\bibinfo
  {journal} {Nat Phys}\ }\textbf {\bibinfo {volume} {11}},\ \bibinfo {pages}
  {421} (\bibinfo {year} {2015})}\BibitemShut {NoStop}%
\bibitem [{\citenamefont {Porer}\ \emph {et~al.}(2014)\citenamefont {Porer},
  \citenamefont {Leierseder}, \citenamefont {M{\'e}nard}, \citenamefont
  {Dachraoui}, \citenamefont {Mouchliadis}, \citenamefont {Perakis},
  \citenamefont {Heinzmann}, \citenamefont {Demsar}, \citenamefont
  {Rossnagel},\ and\ \citenamefont {Huber}}]{Porer:2014aa}%
  \BibitemOpen
  \bibfield  {author} {\bibinfo {author} {\bibfnamefont {M.}~\bibnamefont
  {Porer}}, \bibinfo {author} {\bibfnamefont {U.}~\bibnamefont {Leierseder}},
  \bibinfo {author} {\bibfnamefont {J.~M.}\ \bibnamefont {M{\'e}nard}},
  \bibinfo {author} {\bibfnamefont {H.}~\bibnamefont {Dachraoui}}, \bibinfo
  {author} {\bibfnamefont {L.}~\bibnamefont {Mouchliadis}}, \bibinfo {author}
  {\bibfnamefont {I.~E.}\ \bibnamefont {Perakis}}, \bibinfo {author}
  {\bibfnamefont {U.}~\bibnamefont {Heinzmann}}, \bibinfo {author}
  {\bibfnamefont {J.}~\bibnamefont {Demsar}}, \bibinfo {author} {\bibfnamefont
  {K.}~\bibnamefont {Rossnagel}}, \ and\ \bibinfo {author} {\bibfnamefont
  {R.}~\bibnamefont {Huber}},\ }\href {http://dx.doi.org/10.1038/nmat4042}
  {\bibfield  {journal} {\bibinfo  {journal} {Nat Mater}\ }\textbf {\bibinfo
  {volume} {13}},\ \bibinfo {pages} {857} (\bibinfo {year} {2014})}\BibitemShut
  {NoStop}%
\bibitem [{\citenamefont {Fradkin}\ \emph {et~al.}(2015)\citenamefont
  {Fradkin}, \citenamefont {Kivelson},\ and\ \citenamefont
  {Tranquada}}]{Fradkin:2015aa}%
  \BibitemOpen
  \bibfield  {author} {\bibinfo {author} {\bibfnamefont {E.}~\bibnamefont
  {Fradkin}}, \bibinfo {author} {\bibfnamefont {S.~A.}\ \bibnamefont
  {Kivelson}}, \ and\ \bibinfo {author} {\bibfnamefont {J.~M.}\ \bibnamefont
  {Tranquada}},\ }\href {\doibase 10.1103/RevModPhys.87.457} {\bibfield
  {journal} {\bibinfo  {journal} {Rev. Mod. Phys.}\ }\textbf {\bibinfo {volume}
  {87}},\ \bibinfo {pages} {457} (\bibinfo {year} {2015})}\BibitemShut
  {NoStop}%
\bibitem [{\citenamefont {Aoki}\ \emph {et~al.}(2014)\citenamefont {Aoki},
  \citenamefont {Tsuji}, \citenamefont {Eckstein}, \citenamefont {Kollar},
  \citenamefont {Oka},\ and\ \citenamefont {Werner}}]{Aoki:2014kx}%
  \BibitemOpen
  \bibfield  {author} {\bibinfo {author} {\bibfnamefont {H.}~\bibnamefont
  {Aoki}}, \bibinfo {author} {\bibfnamefont {N.}~\bibnamefont {Tsuji}},
  \bibinfo {author} {\bibfnamefont {M.}~\bibnamefont {Eckstein}}, \bibinfo
  {author} {\bibfnamefont {M.}~\bibnamefont {Kollar}}, \bibinfo {author}
  {\bibfnamefont {T.}~\bibnamefont {Oka}}, \ and\ \bibinfo {author}
  {\bibfnamefont {P.}~\bibnamefont {Werner}},\ }\href {\doibase
  10.1103/RevModPhys.86.779} {\bibfield  {journal} {\bibinfo  {journal} {Rev.
  Mod. Phys.}\ }\textbf {\bibinfo {volume} {86}},\ \bibinfo {pages} {779}
  (\bibinfo {year} {2014})}\BibitemShut {NoStop}%
\bibitem [{\citenamefont {{Kemper}}\ \emph {et~al.}(2016)\citenamefont
  {{Kemper}}, \citenamefont {{Sentef}}, \citenamefont {{Moritz}}, \citenamefont
  {{Devereaux}},\ and\ \citenamefont {{Freericks}}}]{Kemper:2016aa}%
  \BibitemOpen
  \bibfield  {author} {\bibinfo {author} {\bibfnamefont {A.~F.}\ \bibnamefont
  {{Kemper}}}, \bibinfo {author} {\bibfnamefont {M.~A.}\ \bibnamefont
  {{Sentef}}}, \bibinfo {author} {\bibfnamefont {B.}~\bibnamefont {{Moritz}}},
  \bibinfo {author} {\bibfnamefont {T.~P.}\ \bibnamefont {{Devereaux}}}, \ and\
  \bibinfo {author} {\bibfnamefont {J.~K.}\ \bibnamefont {{Freericks}}},\
  }\href@noop {} {\bibfield  {journal} {\bibinfo  {journal} {ArXiv e-prints}\ }
  (\bibinfo {year} {2016})},\ \Eprint {http://arxiv.org/abs/1609.00087}
  {arXiv:1609.00087 [cond-mat.supr-con]} \BibitemShut {NoStop}%
\bibitem [{\citenamefont {Strohmaier}\ \emph {et~al.}(2010)\citenamefont
  {Strohmaier}, \citenamefont {Greif}, \citenamefont {J\"ordens}, \citenamefont
  {Tarruell}, \citenamefont {Moritz}, \citenamefont {Esslinger}, \citenamefont
  {Sensarma}, \citenamefont {Pekker}, \citenamefont {Altman},\ and\
  \citenamefont {Demler}}]{Strohmaier:2010aa}%
  \BibitemOpen
  \bibfield  {author} {\bibinfo {author} {\bibfnamefont {N.}~\bibnamefont
  {Strohmaier}}, \bibinfo {author} {\bibfnamefont {D.}~\bibnamefont {Greif}},
  \bibinfo {author} {\bibfnamefont {R.}~\bibnamefont {J\"ordens}}, \bibinfo
  {author} {\bibfnamefont {L.}~\bibnamefont {Tarruell}}, \bibinfo {author}
  {\bibfnamefont {H.}~\bibnamefont {Moritz}}, \bibinfo {author} {\bibfnamefont
  {T.}~\bibnamefont {Esslinger}}, \bibinfo {author} {\bibfnamefont
  {R.}~\bibnamefont {Sensarma}}, \bibinfo {author} {\bibfnamefont
  {D.}~\bibnamefont {Pekker}}, \bibinfo {author} {\bibfnamefont
  {E.}~\bibnamefont {Altman}}, \ and\ \bibinfo {author} {\bibfnamefont
  {E.}~\bibnamefont {Demler}},\ }\href {\doibase
  10.1103/PhysRevLett.104.080401} {\bibfield  {journal} {\bibinfo  {journal}
  {Phys. Rev. Lett.}\ }\textbf {\bibinfo {volume} {104}},\ \bibinfo {pages}
  {080401} (\bibinfo {year} {2010})}\BibitemShut {NoStop}%
\bibitem [{\citenamefont {Eckstein}\ and\ \citenamefont
  {Werner}(2011)}]{Eckstein:2011aa}%
  \BibitemOpen
  \bibfield  {author} {\bibinfo {author} {\bibfnamefont {M.}~\bibnamefont
  {Eckstein}}\ and\ \bibinfo {author} {\bibfnamefont {P.}~\bibnamefont
  {Werner}},\ }\href {\doibase 10.1103/PhysRevB.84.035122} {\bibfield
  {journal} {\bibinfo  {journal} {Phys. Rev. B}\ }\textbf {\bibinfo {volume}
  {84}},\ \bibinfo {pages} {035122} (\bibinfo {year} {2011})}\BibitemShut
  {NoStop}%
\bibitem [{\citenamefont {Lenar\ifmmode \check{c}\else
  \v{c}\fi{}i\ifmmode~\check{c}\else \v{c}\fi{}}\ and\ \citenamefont
  {Prelov\ifmmode~\check{s}\else
  \v{s}\fi{}ek}(2013)}]{Lenarifmmode-celse-cfiiifmmode-celse-cfi:2013aa}%
  \BibitemOpen
  \bibfield  {author} {\bibinfo {author} {\bibfnamefont {Z.}~\bibnamefont
  {Lenar\ifmmode \check{c}\else \v{c}\fi{}i\ifmmode~\check{c}\else
  \v{c}\fi{}}}\ and\ \bibinfo {author} {\bibfnamefont {P.}~\bibnamefont
  {Prelov\ifmmode~\check{s}\else \v{s}\fi{}ek}},\ }\href {\doibase
  10.1103/PhysRevLett.111.016401} {\bibfield  {journal} {\bibinfo  {journal}
  {Phys. Rev. Lett.}\ }\textbf {\bibinfo {volume} {111}},\ \bibinfo {pages}
  {016401} (\bibinfo {year} {2013})}\BibitemShut {NoStop}%
\bibitem [{\citenamefont {Werner}\ \emph {et~al.}(2014)\citenamefont {Werner},
  \citenamefont {Held},\ and\ \citenamefont {Eckstein}}]{Werner:2014aa}%
  \BibitemOpen
  \bibfield  {author} {\bibinfo {author} {\bibfnamefont {P.}~\bibnamefont
  {Werner}}, \bibinfo {author} {\bibfnamefont {K.}~\bibnamefont {Held}}, \ and\
  \bibinfo {author} {\bibfnamefont {M.}~\bibnamefont {Eckstein}},\ }\href
  {\doibase 10.1103/PhysRevB.90.235102} {\bibfield  {journal} {\bibinfo
  {journal} {Phys. Rev. B}\ }\textbf {\bibinfo {volume} {90}},\ \bibinfo
  {pages} {235102} (\bibinfo {year} {2014})}\BibitemShut {NoStop}%
\bibitem [{\citenamefont {Eckstein}\ and\ \citenamefont
  {Werner}(2013)}]{Eckstein:2013aa}%
  \BibitemOpen
  \bibfield  {author} {\bibinfo {author} {\bibfnamefont {M.}~\bibnamefont
  {Eckstein}}\ and\ \bibinfo {author} {\bibfnamefont {P.}~\bibnamefont
  {Werner}},\ }\href {\doibase 10.1103/PhysRevLett.110.126401} {\bibfield
  {journal} {\bibinfo  {journal} {Phys. Rev. Lett.}\ }\textbf {\bibinfo
  {volume} {110}},\ \bibinfo {pages} {126401} (\bibinfo {year}
  {2013})}\BibitemShut {NoStop}%
\bibitem [{\citenamefont {Werner}\ \emph {et~al.}(2012)\citenamefont {Werner},
  \citenamefont {Tsuji},\ and\ \citenamefont {Eckstein}}]{Werner:2012aa}%
  \BibitemOpen
  \bibfield  {author} {\bibinfo {author} {\bibfnamefont {P.}~\bibnamefont
  {Werner}}, \bibinfo {author} {\bibfnamefont {N.}~\bibnamefont {Tsuji}}, \
  and\ \bibinfo {author} {\bibfnamefont {M.}~\bibnamefont {Eckstein}},\ }\href
  {\doibase 10.1103/PhysRevB.86.205101} {\bibfield  {journal} {\bibinfo
  {journal} {Phys. Rev. B}\ }\textbf {\bibinfo {volume} {86}},\ \bibinfo
  {pages} {205101} (\bibinfo {year} {2012})}\BibitemShut {NoStop}%
\bibitem [{\citenamefont {Gole\ifmmode~\check{z}\else \v{z}\fi{}}\ \emph
  {et~al.}(2015)\citenamefont {Gole\ifmmode~\check{z}\else \v{z}\fi{}},
  \citenamefont {Eckstein},\ and\ \citenamefont
  {Werner}}]{Goleifmmode-zelse-zfi:2015ab}%
  \BibitemOpen
  \bibfield  {author} {\bibinfo {author} {\bibfnamefont {D.}~\bibnamefont
  {Gole\ifmmode~\check{z}\else \v{z}\fi{}}}, \bibinfo {author} {\bibfnamefont
  {M.}~\bibnamefont {Eckstein}}, \ and\ \bibinfo {author} {\bibfnamefont
  {P.}~\bibnamefont {Werner}},\ }\href {\doibase 10.1103/PhysRevB.92.195123}
  {\bibfield  {journal} {\bibinfo  {journal} {Phys. Rev. B}\ }\textbf {\bibinfo
  {volume} {92}},\ \bibinfo {pages} {195123} (\bibinfo {year}
  {2015})}\BibitemShut {NoStop}%
\bibitem [{\citenamefont {Werner}\ and\ \citenamefont
  {Eckstein}(2015)}]{0295-5075-109-3-37002}%
  \BibitemOpen
  \bibfield  {author} {\bibinfo {author} {\bibfnamefont {P.}~\bibnamefont
  {Werner}}\ and\ \bibinfo {author} {\bibfnamefont {M.}~\bibnamefont
  {Eckstein}},\ }\href {http://stacks.iop.org/0295-5075/109/i=3/a=37002}
  {\bibfield  {journal} {\bibinfo  {journal} {EPL (Europhysics Letters)}\
  }\textbf {\bibinfo {volume} {109}},\ \bibinfo {pages} {37002} (\bibinfo
  {year} {2015})}\BibitemShut {NoStop}%
\bibitem [{\citenamefont {Dorfner}\ \emph {et~al.}(2015)\citenamefont
  {Dorfner}, \citenamefont {Vidmar}, \citenamefont {Brockt}, \citenamefont
  {Jeckelmann},\ and\ \citenamefont {Heidrich-Meisner}}]{Dorfner:2015aa}%
  \BibitemOpen
  \bibfield  {author} {\bibinfo {author} {\bibfnamefont {F.}~\bibnamefont
  {Dorfner}}, \bibinfo {author} {\bibfnamefont {L.}~\bibnamefont {Vidmar}},
  \bibinfo {author} {\bibfnamefont {C.}~\bibnamefont {Brockt}}, \bibinfo
  {author} {\bibfnamefont {E.}~\bibnamefont {Jeckelmann}}, \ and\ \bibinfo
  {author} {\bibfnamefont {F.}~\bibnamefont {Heidrich-Meisner}},\ }\href
  {\doibase 10.1103/PhysRevB.91.104302} {\bibfield  {journal} {\bibinfo
  {journal} {Phys. Rev. B}\ }\textbf {\bibinfo {volume} {91}},\ \bibinfo
  {pages} {104302} (\bibinfo {year} {2015})}\BibitemShut {NoStop}%
\bibitem [{\citenamefont {Sayyad}\ and\ \citenamefont
  {Eckstein}(2015)}]{Sayyad:2015aa}%
  \BibitemOpen
  \bibfield  {author} {\bibinfo {author} {\bibfnamefont {S.}~\bibnamefont
  {Sayyad}}\ and\ \bibinfo {author} {\bibfnamefont {M.}~\bibnamefont
  {Eckstein}},\ }\href {\doibase 10.1103/PhysRevB.91.104301} {\bibfield
  {journal} {\bibinfo  {journal} {Phys. Rev. B}\ }\textbf {\bibinfo {volume}
  {91}},\ \bibinfo {pages} {104301} (\bibinfo {year} {2015})}\BibitemShut
  {NoStop}%
\bibitem [{\citenamefont {Murakami}\ \emph {et~al.}(2015)\citenamefont
  {Murakami}, \citenamefont {Werner}, \citenamefont {Tsuji},\ and\
  \citenamefont {Aoki}}]{Murakami:2015aa}%
  \BibitemOpen
  \bibfield  {author} {\bibinfo {author} {\bibfnamefont {Y.}~\bibnamefont
  {Murakami}}, \bibinfo {author} {\bibfnamefont {P.}~\bibnamefont {Werner}},
  \bibinfo {author} {\bibfnamefont {N.}~\bibnamefont {Tsuji}}, \ and\ \bibinfo
  {author} {\bibfnamefont {H.}~\bibnamefont {Aoki}},\ }\href {\doibase
  10.1103/PhysRevB.91.045128} {\bibfield  {journal} {\bibinfo  {journal} {Phys.
  Rev. B}\ }\textbf {\bibinfo {volume} {91}},\ \bibinfo {pages} {045128}
  (\bibinfo {year} {2015})}\BibitemShut {NoStop}%
\bibitem [{\citenamefont {Gole\ifmmode~\check{z}\else \v{z}\fi{}}\ \emph
  {et~al.}(2012)\citenamefont {Gole\ifmmode~\check{z}\else \v{z}\fi{}},
  \citenamefont {Bon\ifmmode~\check{c}\else \v{c}\fi{}a}, \citenamefont
  {Vidmar},\ and\ \citenamefont {Trugman}}]{Goleifmmode-zelse-zfi:2012aa}%
  \BibitemOpen
  \bibfield  {author} {\bibinfo {author} {\bibfnamefont {D.}~\bibnamefont
  {Gole\ifmmode~\check{z}\else \v{z}\fi{}}}, \bibinfo {author} {\bibfnamefont
  {J.}~\bibnamefont {Bon\ifmmode~\check{c}\else \v{c}\fi{}a}}, \bibinfo
  {author} {\bibfnamefont {L.}~\bibnamefont {Vidmar}}, \ and\ \bibinfo {author}
  {\bibfnamefont {S.~A.}\ \bibnamefont {Trugman}},\ }\href {\doibase
  10.1103/PhysRevLett.109.236402} {\bibfield  {journal} {\bibinfo  {journal}
  {Phys. Rev. Lett.}\ }\textbf {\bibinfo {volume} {109}},\ \bibinfo {pages}
  {236402} (\bibinfo {year} {2012})}\BibitemShut {NoStop}%
\bibitem [{\citenamefont {Eckstein}\ and\ \citenamefont
  {Werner}(2016)}]{Eckstein:2016aa}%
  \BibitemOpen
  \bibfield  {author} {\bibinfo {author} {\bibfnamefont {M.}~\bibnamefont
  {Eckstein}}\ and\ \bibinfo {author} {\bibfnamefont {P.}~\bibnamefont
  {Werner}},\ }\href {http://dx.doi.org/10.1038/srep21235} {\bibfield
  {journal} {\bibinfo  {journal} {Scientific Reports}\ }\textbf {\bibinfo
  {volume} {6}},\ \bibinfo {pages} {21235 EP } (\bibinfo {year}
  {2016})}\BibitemShut {NoStop}%
\bibitem [{\citenamefont {Gole\ifmmode~\check{z}\else \v{z}\fi{}}\ \emph
  {et~al.}(2014)\citenamefont {Gole\ifmmode~\check{z}\else \v{z}\fi{}},
  \citenamefont {Bon\ifmmode~\check{c}\else \v{c}\fi{}a}, \citenamefont
  {Mierzejewski},\ and\ \citenamefont {Vidmar}}]{Goleifmmode-zelse-zfi:2014aa}%
  \BibitemOpen
  \bibfield  {author} {\bibinfo {author} {\bibfnamefont {D.}~\bibnamefont
  {Gole\ifmmode~\check{z}\else \v{z}\fi{}}}, \bibinfo {author} {\bibfnamefont
  {J.}~\bibnamefont {Bon\ifmmode~\check{c}\else \v{c}\fi{}a}}, \bibinfo
  {author} {\bibfnamefont {M.}~\bibnamefont {Mierzejewski}}, \ and\ \bibinfo
  {author} {\bibfnamefont {L.}~\bibnamefont {Vidmar}},\ }\href {\doibase
  10.1103/PhysRevB.89.165118} {\bibfield  {journal} {\bibinfo  {journal} {Phys.
  Rev. B}\ }\textbf {\bibinfo {volume} {89}},\ \bibinfo {pages} {165118}
  (\bibinfo {year} {2014})}\BibitemShut {NoStop}%
\bibitem [{\citenamefont {{Golez}}\ \emph {et~al.}(2017)\citenamefont
  {{Golez}}, \citenamefont {{Boehnke}}, \citenamefont {{Strand}}, \citenamefont
  {{Eckstein}},\ and\ \citenamefont {{Werner}}}]{Golez:2017aa}%
  \BibitemOpen
  \bibfield  {author} {\bibinfo {author} {\bibfnamefont {D.}~\bibnamefont
  {{Golez}}}, \bibinfo {author} {\bibfnamefont {L.}~\bibnamefont {{Boehnke}}},
  \bibinfo {author} {\bibfnamefont {H.}~\bibnamefont {{Strand}}}, \bibinfo
  {author} {\bibfnamefont {M.}~\bibnamefont {{Eckstein}}}, \ and\ \bibinfo
  {author} {\bibfnamefont {P.}~\bibnamefont {{Werner}}},\ }\href@noop {}
  {\bibfield  {journal} {\bibinfo  {journal} {ArXiv e-prints}\ } (\bibinfo
  {year} {2017})},\ \Eprint {http://arxiv.org/abs/1702.04952} {arXiv:1702.04952
  [cond-mat.str-el]} \BibitemShut {NoStop}%
\bibitem [{\citenamefont {Werner}\ and\ \citenamefont
  {Eckstein}(2016)}]{doi:10.1063/1.4935245}%
  \BibitemOpen
  \bibfield  {author} {\bibinfo {author} {\bibfnamefont {P.}~\bibnamefont
  {Werner}}\ and\ \bibinfo {author} {\bibfnamefont {M.}~\bibnamefont
  {Eckstein}},\ }\href {\doibase 10.1063/1.4935245} {\bibfield  {journal}
  {\bibinfo  {journal} {Structural Dynamics}\ }\textbf {\bibinfo {volume}
  {3}},\ \bibinfo {pages} {023603} (\bibinfo {year} {2016})}\BibitemShut
  {NoStop}%
\bibitem [{\citenamefont {Oelsen}\ \emph {et~al.}(2011)\citenamefont {Oelsen},
  \citenamefont {Seibold},\ and\ \citenamefont {B\"unemann}}]{Oelsen:2011cr}%
  \BibitemOpen
  \bibfield  {author} {\bibinfo {author} {\bibfnamefont {E.~v.}\ \bibnamefont
  {Oelsen}}, \bibinfo {author} {\bibfnamefont {G.}~\bibnamefont {Seibold}}, \
  and\ \bibinfo {author} {\bibfnamefont {J.}~\bibnamefont {B\"unemann}},\
  }\href {\doibase 10.1103/PhysRevLett.107.076402} {\bibfield  {journal}
  {\bibinfo  {journal} {Phys. Rev. Lett.}\ }\textbf {\bibinfo {volume} {107}},\
  \bibinfo {pages} {076402} (\bibinfo {year} {2011})}\BibitemShut {NoStop}%
\bibitem [{\citenamefont {von Oelsen}\ \emph {et~al.}(2011)\citenamefont {von
  Oelsen}, \citenamefont {Seibold},\ and\ \citenamefont
  {B{\"u}nemann}}]{Oelsen:2011nx}%
  \BibitemOpen
  \bibfield  {author} {\bibinfo {author} {\bibfnamefont {E.}~\bibnamefont {von
  Oelsen}}, \bibinfo {author} {\bibfnamefont {G.}~\bibnamefont {Seibold}}, \
  and\ \bibinfo {author} {\bibfnamefont {J.}~\bibnamefont {B{\"u}nemann}},\
  }\href {http://stacks.iop.org/1367-2630/13/i=11/a=113031} {\bibfield
  {journal} {\bibinfo  {journal} {New Journal of Physics}\ }\textbf {\bibinfo
  {volume} {13}} (\bibinfo {year} {2011})}\BibitemShut {NoStop}%
\bibitem [{\citenamefont {Behrmann}\ \emph {et~al.}(2013)\citenamefont
  {Behrmann}, \citenamefont {Fabrizio},\ and\ \citenamefont
  {Lechermann}}]{Behrmann:2013aa}%
  \BibitemOpen
  \bibfield  {author} {\bibinfo {author} {\bibfnamefont {M.}~\bibnamefont
  {Behrmann}}, \bibinfo {author} {\bibfnamefont {M.}~\bibnamefont {Fabrizio}},
  \ and\ \bibinfo {author} {\bibfnamefont {F.}~\bibnamefont {Lechermann}},\
  }\href {\doibase 10.1103/PhysRevB.88.035116} {\bibfield  {journal} {\bibinfo
  {journal} {Phys. Rev. B}\ }\textbf {\bibinfo {volume} {88}},\ \bibinfo
  {pages} {035116} (\bibinfo {year} {2013})}\BibitemShut {NoStop}%
\bibitem [{\citenamefont {Behrmann}\ and\ \citenamefont
  {Lechermann}(2015)}]{Behrmann:2015aa}%
  \BibitemOpen
  \bibfield  {author} {\bibinfo {author} {\bibfnamefont {M.}~\bibnamefont
  {Behrmann}}\ and\ \bibinfo {author} {\bibfnamefont {F.}~\bibnamefont
  {Lechermann}},\ }\href {\doibase 10.1103/PhysRevB.91.075110} {\bibfield
  {journal} {\bibinfo  {journal} {Phys. Rev. B}\ }\textbf {\bibinfo {volume}
  {91}},\ \bibinfo {pages} {075110} (\bibinfo {year} {2015})}\BibitemShut
  {NoStop}%
\bibitem [{\citenamefont {Sandri}\ and\ \citenamefont
  {Fabrizio}(2015)}]{Sandri:2015aa}%
  \BibitemOpen
  \bibfield  {author} {\bibinfo {author} {\bibfnamefont {M.}~\bibnamefont
  {Sandri}}\ and\ \bibinfo {author} {\bibfnamefont {M.}~\bibnamefont
  {Fabrizio}},\ }\href {\doibase 10.1103/PhysRevB.91.115102} {\bibfield
  {journal} {\bibinfo  {journal} {Phys. Rev. B}\ }\textbf {\bibinfo {volume}
  {91}},\ \bibinfo {pages} {115102} (\bibinfo {year} {2015})}\BibitemShut
  {NoStop}%
\bibitem [{\citenamefont {Behrmann}\ \emph {et~al.}(2016)\citenamefont
  {Behrmann}, \citenamefont {Lichtenstein}, \citenamefont {Katsnelson},\ and\
  \citenamefont {Lechermann}}]{Behrmann:2016aa}%
  \BibitemOpen
  \bibfield  {author} {\bibinfo {author} {\bibfnamefont {M.}~\bibnamefont
  {Behrmann}}, \bibinfo {author} {\bibfnamefont {A.~I.}\ \bibnamefont
  {Lichtenstein}}, \bibinfo {author} {\bibfnamefont {M.~I.}\ \bibnamefont
  {Katsnelson}}, \ and\ \bibinfo {author} {\bibfnamefont {F.}~\bibnamefont
  {Lechermann}},\ }\href {\doibase 10.1103/PhysRevB.94.165120} {\bibfield
  {journal} {\bibinfo  {journal} {Phys. Rev. B}\ }\textbf {\bibinfo {volume}
  {94}},\ \bibinfo {pages} {165120} (\bibinfo {year} {2016})}\BibitemShut
  {NoStop}%
\bibitem [{\citenamefont {{Mazza}}\ and\ \citenamefont
  {{Georges}}(2017)}]{Mazza:2017aa}%
  \BibitemOpen
  \bibfield  {author} {\bibinfo {author} {\bibfnamefont {G.}~\bibnamefont
  {{Mazza}}}\ and\ \bibinfo {author} {\bibfnamefont {A.}~\bibnamefont
  {{Georges}}},\ }\href@noop {} {\bibfield  {journal} {\bibinfo  {journal}
  {ArXiv e-prints}\ } (\bibinfo {year} {2017})},\ \Eprint
  {http://arxiv.org/abs/1702.04675} {arXiv:1702.04675 [cond-mat.supr-con]}
  \BibitemShut {NoStop}%
\bibitem [{\citenamefont {{Sutter}}\ \emph {et~al.}(2016)\citenamefont
  {{Sutter}}, \citenamefont {{Fatuzzo}}, \citenamefont {{Moser}}, \citenamefont
  {{Kim}}, \citenamefont {{Fittipaldi}}, \citenamefont {{Vecchione}},
  \citenamefont {{Granata}}, \citenamefont {{Sassa}}, \citenamefont
  {{Cossalter}}, \citenamefont {{Gatti}}, \citenamefont {{Grioni}},
  \citenamefont {{Ronnow}}, \citenamefont {{Plumb}}, \citenamefont {{Matt}},
  \citenamefont {{Shi}}, \citenamefont {{Hoesch}}, \citenamefont {{Kim}},
  \citenamefont {{Chang}}, \citenamefont {{Jeng}}, \citenamefont {{Jozwiak}},
  \citenamefont {{Bostwick}}, \citenamefont {{Rotenberg}}, \citenamefont
  {{Georges}}, \citenamefont {{Neupert}},\ and\ \citenamefont
  {{Chang}}}]{Sutter:2016aa}%
  \BibitemOpen
  \bibfield  {author} {\bibinfo {author} {\bibfnamefont {D.}~\bibnamefont
  {{Sutter}}}, \bibinfo {author} {\bibfnamefont {C.~G.}\ \bibnamefont
  {{Fatuzzo}}}, \bibinfo {author} {\bibfnamefont {S.}~\bibnamefont {{Moser}}},
  \bibinfo {author} {\bibfnamefont {M.}~\bibnamefont {{Kim}}}, \bibinfo
  {author} {\bibfnamefont {R.}~\bibnamefont {{Fittipaldi}}}, \bibinfo {author}
  {\bibfnamefont {A.}~\bibnamefont {{Vecchione}}}, \bibinfo {author}
  {\bibfnamefont {V.}~\bibnamefont {{Granata}}}, \bibinfo {author}
  {\bibfnamefont {Y.}~\bibnamefont {{Sassa}}}, \bibinfo {author} {\bibfnamefont
  {F.}~\bibnamefont {{Cossalter}}}, \bibinfo {author} {\bibfnamefont
  {G.}~\bibnamefont {{Gatti}}}, \bibinfo {author} {\bibfnamefont
  {M.}~\bibnamefont {{Grioni}}}, \bibinfo {author} {\bibfnamefont {H.~M.}\
  \bibnamefont {{Ronnow}}}, \bibinfo {author} {\bibfnamefont {N.~C.}\
  \bibnamefont {{Plumb}}}, \bibinfo {author} {\bibfnamefont {C.~E.}\
  \bibnamefont {{Matt}}}, \bibinfo {author} {\bibfnamefont {M.}~\bibnamefont
  {{Shi}}}, \bibinfo {author} {\bibfnamefont {M.}~\bibnamefont {{Hoesch}}},
  \bibinfo {author} {\bibfnamefont {T.~K.}\ \bibnamefont {{Kim}}}, \bibinfo
  {author} {\bibfnamefont {T.~R.}\ \bibnamefont {{Chang}}}, \bibinfo {author}
  {\bibfnamefont {H.~T.}\ \bibnamefont {{Jeng}}}, \bibinfo {author}
  {\bibfnamefont {C.}~\bibnamefont {{Jozwiak}}}, \bibinfo {author}
  {\bibfnamefont {A.}~\bibnamefont {{Bostwick}}}, \bibinfo {author}
  {\bibfnamefont {E.}~\bibnamefont {{Rotenberg}}}, \bibinfo {author}
  {\bibfnamefont {A.}~\bibnamefont {{Georges}}}, \bibinfo {author}
  {\bibfnamefont {T.}~\bibnamefont {{Neupert}}}, \ and\ \bibinfo {author}
  {\bibfnamefont {J.}~\bibnamefont {{Chang}}},\ }\href@noop {} {\bibfield
  {journal} {\bibinfo  {journal} {ArXiv e-prints}\ } (\bibinfo {year}
  {2016})},\ \Eprint {http://arxiv.org/abs/1610.02854} {arXiv:1610.02854
  [cond-mat.str-el]} \BibitemShut {NoStop}%
\bibitem [{\citenamefont {Sugano}\ \emph {et~al.}(1970)\citenamefont {Sugano},
  \citenamefont {Tanabe},\ and\ \citenamefont {Kamimura}}]{Sugaon:1970kx}%
  \BibitemOpen
  \bibfield  {author} {\bibinfo {author} {\bibfnamefont {S.}~\bibnamefont
  {Sugano}}, \bibinfo {author} {\bibfnamefont {Y.}~\bibnamefont {Tanabe}}, \
  and\ \bibinfo {author} {\bibfnamefont {H.}~\bibnamefont {Kamimura}},\
  }\href@noop {} {\emph {\bibinfo {title} {Multiplets of Transition-Metal Ions
  in Crystals}}}\ (\bibinfo  {publisher} {Academic Press Inc.},\ \bibinfo
  {year} {1970})\BibitemShut {NoStop}%
\bibitem [{\citenamefont {Kanamori}(1963)}]{Kanamori:1963aa}%
  \BibitemOpen
  \bibfield  {author} {\bibinfo {author} {\bibfnamefont {J.}~\bibnamefont
  {Kanamori}},\ }\href {\doibase 10.1143/PTP.30.275} {\bibfield  {journal}
  {\bibinfo  {journal} {Prog. Theor. Phys.}\ }\textbf {\bibinfo {volume}
  {30}},\ \bibinfo {pages} {275} (\bibinfo {year} {1963})}\BibitemShut
  {NoStop}%
\bibitem [{\citenamefont {Hoshino}\ and\ \citenamefont
  {Werner}(2016)}]{Hoshino:2016aa}%
  \BibitemOpen
  \bibfield  {author} {\bibinfo {author} {\bibfnamefont {S.}~\bibnamefont
  {Hoshino}}\ and\ \bibinfo {author} {\bibfnamefont {P.}~\bibnamefont
  {Werner}},\ }\href {\doibase 10.1103/PhysRevB.93.155161} {\bibfield
  {journal} {\bibinfo  {journal} {Phys. Rev. B}\ }\textbf {\bibinfo {volume}
  {93}},\ \bibinfo {pages} {155161} (\bibinfo {year} {2016})}\BibitemShut
  {NoStop}%
\bibitem [{\citenamefont {{Steiner}}\ \emph {et~al.}(2016)\citenamefont
  {{Steiner}}, \citenamefont {{Hoshino}}, \citenamefont {{Nomura}},\ and\
  \citenamefont {{Werner}}}]{Steiner:2016aa}%
  \BibitemOpen
  \bibfield  {author} {\bibinfo {author} {\bibfnamefont {K.}~\bibnamefont
  {{Steiner}}}, \bibinfo {author} {\bibfnamefont {S.}~\bibnamefont
  {{Hoshino}}}, \bibinfo {author} {\bibfnamefont {Y.}~\bibnamefont {{Nomura}}},
  \ and\ \bibinfo {author} {\bibfnamefont {P.}~\bibnamefont {{Werner}}},\
  }\href@noop {} {\bibfield  {journal} {\bibinfo  {journal} {arXiv:1605.06410}\
  } (\bibinfo {year} {2016})},\ \Eprint {http://arxiv.org/abs/1605.06410}
  {arXiv:1605.06410 [cond-mat.str-el]} \BibitemShut {NoStop}%
\bibitem [{\citenamefont {Tsuji}\ and\ \citenamefont
  {Werner}(2013)}]{Tsuji:2013mi}%
  \BibitemOpen
  \bibfield  {author} {\bibinfo {author} {\bibfnamefont {N.}~\bibnamefont
  {Tsuji}}\ and\ \bibinfo {author} {\bibfnamefont {P.}~\bibnamefont {Werner}},\
  }\href {\doibase 10.1103/PhysRevB.88.165115} {\bibfield  {journal} {\bibinfo
  {journal} {Phys. Rev. B}\ }\textbf {\bibinfo {volume} {88}},\ \bibinfo
  {pages} {165115} (\bibinfo {year} {2013})}\BibitemShut {NoStop}%
\bibitem [{\citenamefont {Tsuji}\ \emph {et~al.}(2013)\citenamefont {Tsuji},
  \citenamefont {Eckstein},\ and\ \citenamefont {Werner}}]{Tsuji:2013kkk}%
  \BibitemOpen
  \bibfield  {author} {\bibinfo {author} {\bibfnamefont {N.}~\bibnamefont
  {Tsuji}}, \bibinfo {author} {\bibfnamefont {M.}~\bibnamefont {Eckstein}}, \
  and\ \bibinfo {author} {\bibfnamefont {P.}~\bibnamefont {Werner}},\ }\href
  {\doibase 10.1103/PhysRevLett.110.136404} {\bibfield  {journal} {\bibinfo
  {journal} {Phys. Rev. Lett.}\ }\textbf {\bibinfo {volume} {110}},\ \bibinfo
  {pages} {136404} (\bibinfo {year} {2013})}\BibitemShut {NoStop}%
\bibitem [{\citenamefont {Eckstein}\ and\ \citenamefont
  {Werner}(2010)}]{Eckstein:2010fk}%
  \BibitemOpen
  \bibfield  {author} {\bibinfo {author} {\bibfnamefont {M.}~\bibnamefont
  {Eckstein}}\ and\ \bibinfo {author} {\bibfnamefont {P.}~\bibnamefont
  {Werner}},\ }\href {\doibase 10.1103/PhysRevB.82.115115} {\bibfield
  {journal} {\bibinfo  {journal} {Phys. Rev. B}\ }\textbf {\bibinfo {volume}
  {82}},\ \bibinfo {pages} {115115} (\bibinfo {year} {2010})}\BibitemShut
  {NoStop}%
\bibitem [{\citenamefont {Baym}\ and\ \citenamefont
  {Kadanoff}(1961)}]{Baym:1961tw}%
  \BibitemOpen
  \bibfield  {author} {\bibinfo {author} {\bibfnamefont {G.}~\bibnamefont
  {Baym}}\ and\ \bibinfo {author} {\bibfnamefont {L.~P.}\ \bibnamefont
  {Kadanoff}},\ }\href {\doibase 10.1103/PhysRev.124.287} {\bibfield  {journal}
  {\bibinfo  {journal} {Phys. Rev.}\ }\textbf {\bibinfo {volume} {124}},\
  \bibinfo {pages} {287} (\bibinfo {year} {1961})}\BibitemShut {NoStop}%
\bibitem [{\citenamefont {Baym}(1962)}]{Baym:1962qo}%
  \BibitemOpen
  \bibfield  {author} {\bibinfo {author} {\bibfnamefont {G.}~\bibnamefont
  {Baym}},\ }\href {\doibase 10.1103/PhysRev.127.1391} {\bibfield  {journal}
  {\bibinfo  {journal} {Phys. Rev.}\ }\textbf {\bibinfo {volume} {127}},\
  \bibinfo {pages} {1391} (\bibinfo {year} {1962})}\BibitemShut {NoStop}%
\bibitem [{\citenamefont {Koga}\ \emph {et~al.}(2005)\citenamefont {Koga},
  \citenamefont {Inaba},\ and\ \citenamefont {Kawakami}}]{Koga:2005aa}%
  \BibitemOpen
  \bibfield  {author} {\bibinfo {author} {\bibfnamefont {A.}~\bibnamefont
  {Koga}}, \bibinfo {author} {\bibfnamefont {K.}~\bibnamefont {Inaba}}, \ and\
  \bibinfo {author} {\bibfnamefont {N.}~\bibnamefont {Kawakami}},\ }\href
  {http://dx.doi.org/10.1143/PTPS.160.253} {\bibfield  {journal} {\bibinfo
  {journal} {Prog. Theor. Phys.}\ }\textbf {\bibinfo {volume} {160}},\ \bibinfo
  {pages} {253} (\bibinfo {year} {2005})}\BibitemShut {NoStop}%
\bibitem [{\citenamefont {Pruschke}\ and\ \citenamefont
  {Bulla}(2005)}]{Pruschke:2005aa}%
  \BibitemOpen
  \bibfield  {author} {\bibinfo {author} {\bibfnamefont {T.}~\bibnamefont
  {Pruschke}}\ and\ \bibinfo {author} {\bibfnamefont {R.}~\bibnamefont
  {Bulla}},\ }\href {\doibase 10.1140/epjb/e2005-00117-4} {\bibfield  {journal}
  {\bibinfo  {journal} {Eur. Phys. J. B}\ }\textbf {\bibinfo {volume} {44}},\
  \bibinfo {pages} {217} (\bibinfo {year} {2005})}\BibitemShut {NoStop}%
\bibitem [{\citenamefont {Werner}\ and\ \citenamefont
  {Millis}(2006)}]{Werner:2006qy}%
  \BibitemOpen
  \bibfield  {author} {\bibinfo {author} {\bibfnamefont {P.}~\bibnamefont
  {Werner}}\ and\ \bibinfo {author} {\bibfnamefont {A.~J.}\ \bibnamefont
  {Millis}},\ }\href {\doibase 10.1103/PhysRevB.74.155107} {\bibfield
  {journal} {\bibinfo  {journal} {Phys. Rev. B}\ }\textbf {\bibinfo {volume}
  {74}},\ \bibinfo {pages} {155107} (\bibinfo {year} {2006})}\BibitemShut
  {NoStop}%
\bibitem [{\citenamefont {Sakai}\ \emph {et~al.}(2006)\citenamefont {Sakai},
  \citenamefont {Arita}, \citenamefont {Held},\ and\ \citenamefont
  {Aoki}}]{Sakai:2006aa}%
  \BibitemOpen
  \bibfield  {author} {\bibinfo {author} {\bibfnamefont {S.}~\bibnamefont
  {Sakai}}, \bibinfo {author} {\bibfnamefont {R.}~\bibnamefont {Arita}},
  \bibinfo {author} {\bibfnamefont {K.}~\bibnamefont {Held}}, \ and\ \bibinfo
  {author} {\bibfnamefont {H.}~\bibnamefont {Aoki}},\ }\href {\doibase
  10.1103/PhysRevB.74.155102} {\bibfield  {journal} {\bibinfo  {journal} {Phys.
  Rev. B}\ }\textbf {\bibinfo {volume} {74}},\ \bibinfo {pages} {155102}
  (\bibinfo {year} {2006})}\BibitemShut {NoStop}%
\bibitem [{\citenamefont {Werner}\ and\ \citenamefont
  {Millis}(2007)}]{Werner:2007lr}%
  \BibitemOpen
  \bibfield  {author} {\bibinfo {author} {\bibfnamefont {P.}~\bibnamefont
  {Werner}}\ and\ \bibinfo {author} {\bibfnamefont {A.~J.}\ \bibnamefont
  {Millis}},\ }\href {\doibase 10.1103/PhysRevLett.99.126405} {\bibfield
  {journal} {\bibinfo  {journal} {Phys. Rev. Lett.}\ }\textbf {\bibinfo
  {volume} {99}},\ \bibinfo {pages} {126405} (\bibinfo {year}
  {2007})}\BibitemShut {NoStop}%
\bibitem [{\citenamefont {Bulla}\ \emph {et~al.}(2008)\citenamefont {Bulla},
  \citenamefont {Costi},\ and\ \citenamefont {Pruschke}}]{Bulla:2008kx}%
  \BibitemOpen
  \bibfield  {author} {\bibinfo {author} {\bibfnamefont {R.}~\bibnamefont
  {Bulla}}, \bibinfo {author} {\bibfnamefont {T.~A.}\ \bibnamefont {Costi}}, \
  and\ \bibinfo {author} {\bibfnamefont {T.}~\bibnamefont {Pruschke}},\ }\href
  {\doibase 10.1103/RevModPhys.80.395} {\bibfield  {journal} {\bibinfo
  {journal} {Rev. Mod. Phys.}\ }\textbf {\bibinfo {volume} {80}},\ \bibinfo
  {pages} {395} (\bibinfo {year} {2008})}\BibitemShut {NoStop}%
\bibitem [{\citenamefont {Peters}\ and\ \citenamefont
  {Pruschke}(2010)}]{Peters:2010aa}%
  \BibitemOpen
  \bibfield  {author} {\bibinfo {author} {\bibfnamefont {R.}~\bibnamefont
  {Peters}}\ and\ \bibinfo {author} {\bibfnamefont {T.}~\bibnamefont
  {Pruschke}},\ }\href {\doibase 10.1103/PhysRevB.81.035112} {\bibfield
  {journal} {\bibinfo  {journal} {Phys. Rev. B}\ }\textbf {\bibinfo {volume}
  {81}},\ \bibinfo {pages} {035112} (\bibinfo {year} {2010})}\BibitemShut
  {NoStop}%
\bibitem [{\citenamefont {Peters}\ \emph {et~al.}(2011)\citenamefont {Peters},
  \citenamefont {Kawakami},\ and\ \citenamefont {Pruschke}}]{Peters:2011aa}%
  \BibitemOpen
  \bibfield  {author} {\bibinfo {author} {\bibfnamefont {R.}~\bibnamefont
  {Peters}}, \bibinfo {author} {\bibfnamefont {N.}~\bibnamefont {Kawakami}}, \
  and\ \bibinfo {author} {\bibfnamefont {T.}~\bibnamefont {Pruschke}},\ }\href
  {\doibase 10.1103/PhysRevB.83.125110} {\bibfield  {journal} {\bibinfo
  {journal} {Phys. Rev. B}\ }\textbf {\bibinfo {volume} {83}},\ \bibinfo
  {pages} {125110} (\bibinfo {year} {2011})}\BibitemShut {NoStop}%
\bibitem [{\citenamefont {Antipov}\ \emph {et~al.}(2012)\citenamefont
  {Antipov}, \citenamefont {Krivenko}, \citenamefont {Anisimov}, \citenamefont
  {Lichtenstein},\ and\ \citenamefont {Rubtsov}}]{Antipov:2012aa}%
  \BibitemOpen
  \bibfield  {author} {\bibinfo {author} {\bibfnamefont {A.~E.}\ \bibnamefont
  {Antipov}}, \bibinfo {author} {\bibfnamefont {I.~S.}\ \bibnamefont
  {Krivenko}}, \bibinfo {author} {\bibfnamefont {V.~I.}\ \bibnamefont
  {Anisimov}}, \bibinfo {author} {\bibfnamefont {A.~I.}\ \bibnamefont
  {Lichtenstein}}, \ and\ \bibinfo {author} {\bibfnamefont {A.~N.}\
  \bibnamefont {Rubtsov}},\ }\href {\doibase 10.1103/PhysRevB.86.155107}
  {\bibfield  {journal} {\bibinfo  {journal} {Phys. Rev. B}\ }\textbf {\bibinfo
  {volume} {86}},\ \bibinfo {pages} {155107} (\bibinfo {year}
  {2012})}\BibitemShut {NoStop}%
\bibitem [{\citenamefont {Werner}\ \emph {et~al.}(2006)\citenamefont {Werner},
  \citenamefont {Comanac}, \citenamefont {de' Medici}, \citenamefont {Troyer},\
  and\ \citenamefont {Millis}}]{Werner:2006rt}%
  \BibitemOpen
  \bibfield  {author} {\bibinfo {author} {\bibfnamefont {P.}~\bibnamefont
  {Werner}}, \bibinfo {author} {\bibfnamefont {A.}~\bibnamefont {Comanac}},
  \bibinfo {author} {\bibfnamefont {L.}~\bibnamefont {de' Medici}}, \bibinfo
  {author} {\bibfnamefont {M.}~\bibnamefont {Troyer}}, \ and\ \bibinfo {author}
  {\bibfnamefont {A.~J.}\ \bibnamefont {Millis}},\ }\href {\doibase
  10.1103/PhysRevLett.97.076405} {\bibfield  {journal} {\bibinfo  {journal}
  {Phys. Rev. Lett.}\ }\textbf {\bibinfo {volume} {97}},\ \bibinfo {pages}
  {076405} (\bibinfo {year} {2006})}\BibitemShut {NoStop}%
\bibitem [{\citenamefont {Gull}\ \emph {et~al.}(2011)\citenamefont {Gull},
  \citenamefont {Millis}, \citenamefont {Lichtenstein}, \citenamefont
  {Rubtsov}, \citenamefont {Troyer},\ and\ \citenamefont
  {Werner}}]{Gull:2011lr}%
  \BibitemOpen
  \bibfield  {author} {\bibinfo {author} {\bibfnamefont {E.}~\bibnamefont
  {Gull}}, \bibinfo {author} {\bibfnamefont {A.~J.}\ \bibnamefont {Millis}},
  \bibinfo {author} {\bibfnamefont {A.~I.}\ \bibnamefont {Lichtenstein}},
  \bibinfo {author} {\bibfnamefont {A.~N.}\ \bibnamefont {Rubtsov}}, \bibinfo
  {author} {\bibfnamefont {M.}~\bibnamefont {Troyer}}, \ and\ \bibinfo {author}
  {\bibfnamefont {P.}~\bibnamefont {Werner}},\ }\href {\doibase
  10.1103/RevModPhys.83.349} {\bibfield  {journal} {\bibinfo  {journal} {Rev.
  Mod. Phys.}\ }\textbf {\bibinfo {volume} {83}},\ \bibinfo {pages} {349}
  (\bibinfo {year} {2011})}\BibitemShut {NoStop}%
\bibitem [{\citenamefont {Seth}\ \emph {et~al.}(2016)\citenamefont {Seth},
  \citenamefont {Krivenko}, \citenamefont {Ferrero},\ and\ \citenamefont
  {Parcollet}}]{Seth2016274}%
  \BibitemOpen
  \bibfield  {author} {\bibinfo {author} {\bibfnamefont {P.}~\bibnamefont
  {Seth}}, \bibinfo {author} {\bibfnamefont {I.}~\bibnamefont {Krivenko}},
  \bibinfo {author} {\bibfnamefont {M.}~\bibnamefont {Ferrero}}, \ and\
  \bibinfo {author} {\bibfnamefont {O.}~\bibnamefont {Parcollet}},\ }\href
  {\doibase http://dx.doi.org/10.1016/j.cpc.2015.10.023} {\bibfield  {journal}
  {\bibinfo  {journal} {Computer Physics Communications}\ }\textbf {\bibinfo
  {volume} {200}},\ \bibinfo {pages} {274 } (\bibinfo {year}
  {2016})}\BibitemShut {NoStop}%
\bibitem [{\citenamefont {Parcollet}\ \emph {et~al.}(2015)\citenamefont
  {Parcollet}, \citenamefont {Ferrero}, \citenamefont {Ayral}, \citenamefont
  {Hafermann}, \citenamefont {Krivenko}, \citenamefont {Messio},\ and\
  \citenamefont {Seth}}]{Parcollet2015398}%
  \BibitemOpen
  \bibfield  {author} {\bibinfo {author} {\bibfnamefont {O.}~\bibnamefont
  {Parcollet}}, \bibinfo {author} {\bibfnamefont {M.}~\bibnamefont {Ferrero}},
  \bibinfo {author} {\bibfnamefont {T.}~\bibnamefont {Ayral}}, \bibinfo
  {author} {\bibfnamefont {H.}~\bibnamefont {Hafermann}}, \bibinfo {author}
  {\bibfnamefont {I.}~\bibnamefont {Krivenko}}, \bibinfo {author}
  {\bibfnamefont {L.}~\bibnamefont {Messio}}, \ and\ \bibinfo {author}
  {\bibfnamefont {P.}~\bibnamefont {Seth}},\ }\href {\doibase
  http://dx.doi.org/10.1016/j.cpc.2015.04.023} {\bibfield  {journal} {\bibinfo
  {journal} {Computer Physics Communications}\ }\textbf {\bibinfo {volume}
  {196}},\ \bibinfo {pages} {398 } (\bibinfo {year} {2015})}\BibitemShut
  {NoStop}%
\bibitem [{\citenamefont {Boehnke}\ \emph {et~al.}(2011)\citenamefont
  {Boehnke}, \citenamefont {Hafermann}, \citenamefont {Ferrero}, \citenamefont
  {Lechermann},\ and\ \citenamefont {Parcollet}}]{Boehnke:2011fk}%
  \BibitemOpen
  \bibfield  {author} {\bibinfo {author} {\bibfnamefont {L.}~\bibnamefont
  {Boehnke}}, \bibinfo {author} {\bibfnamefont {H.}~\bibnamefont {Hafermann}},
  \bibinfo {author} {\bibfnamefont {M.}~\bibnamefont {Ferrero}}, \bibinfo
  {author} {\bibfnamefont {F.}~\bibnamefont {Lechermann}}, \ and\ \bibinfo
  {author} {\bibfnamefont {O.}~\bibnamefont {Parcollet}},\ }\href {\doibase
  10.1103/PhysRevB.84.075145} {\bibfield  {journal} {\bibinfo  {journal} {Phys.
  Rev. B}\ }\textbf {\bibinfo {volume} {84}},\ \bibinfo {pages} {075145}
  (\bibinfo {year} {2011})}\BibitemShut {NoStop}%
\bibitem [{\citenamefont {Strand}\ \emph {et~al.}(2015)\citenamefont {Strand},
  \citenamefont {Eckstein},\ and\ \citenamefont {Werner}}]{Strand:2015ac}%
  \BibitemOpen
  \bibfield  {author} {\bibinfo {author} {\bibfnamefont {H.~U.~R.}\
  \bibnamefont {Strand}}, \bibinfo {author} {\bibfnamefont {M.}~\bibnamefont
  {Eckstein}}, \ and\ \bibinfo {author} {\bibfnamefont {P.}~\bibnamefont
  {Werner}},\ }\href {\doibase 10.1103/PhysRevA.92.063602} {\bibfield
  {journal} {\bibinfo  {journal} {Phys. Rev. A}\ }\textbf {\bibinfo {volume}
  {92}},\ \bibinfo {pages} {063602} (\bibinfo {year} {2015})}\BibitemShut
  {NoStop}%
\bibitem [{\citenamefont {Aub\"{o}ck}\ \emph {et~al.}(2012)\citenamefont
  {Aub\"{o}ck}, \citenamefont {Consani}, \citenamefont {van Mourik},\ and\
  \citenamefont {Chergui}}]{Aubock:12}%
  \BibitemOpen
  \bibfield  {author} {\bibinfo {author} {\bibfnamefont {G.}~\bibnamefont
  {Aub\"{o}ck}}, \bibinfo {author} {\bibfnamefont {C.}~\bibnamefont {Consani}},
  \bibinfo {author} {\bibfnamefont {F.}~\bibnamefont {van Mourik}}, \ and\
  \bibinfo {author} {\bibfnamefont {M.}~\bibnamefont {Chergui}},\ }\href
  {\doibase 10.1364/OL.37.002337} {\bibfield  {journal} {\bibinfo  {journal}
  {Opt. Lett.}\ }\textbf {\bibinfo {volume} {37}},\ \bibinfo {pages} {2337}
  (\bibinfo {year} {2012})}\BibitemShut {NoStop}%
\bibitem [{\citenamefont {{Baldini}}\ \emph {et~al.}(2016)\citenamefont
  {{Baldini}}, \citenamefont {{Chiodo}}, \citenamefont {{Dominguez}},
  \citenamefont {{Palummo}}, \citenamefont {{Moser}}, \citenamefont {{Yazdi}},
  \citenamefont {{Aub{\"o}ck}}, \citenamefont {{Mallett}}, \citenamefont
  {{Berger}}, \citenamefont {{Magrez}}, \citenamefont {{Bernhard}},
  \citenamefont {{Grioni}}, \citenamefont {{Rubio}},\ and\ \citenamefont
  {{Chergui}}}]{Baldini:2016aa}%
  \BibitemOpen
  \bibfield  {author} {\bibinfo {author} {\bibfnamefont {E.}~\bibnamefont
  {{Baldini}}}, \bibinfo {author} {\bibfnamefont {L.}~\bibnamefont {{Chiodo}}},
  \bibinfo {author} {\bibfnamefont {A.}~\bibnamefont {{Dominguez}}}, \bibinfo
  {author} {\bibfnamefont {M.}~\bibnamefont {{Palummo}}}, \bibinfo {author}
  {\bibfnamefont {S.}~\bibnamefont {{Moser}}}, \bibinfo {author} {\bibfnamefont
  {M.}~\bibnamefont {{Yazdi}}}, \bibinfo {author} {\bibfnamefont
  {G.}~\bibnamefont {{Aub{\"o}ck}}}, \bibinfo {author} {\bibfnamefont
  {B.~P.~P.}\ \bibnamefont {{Mallett}}}, \bibinfo {author} {\bibfnamefont
  {H.}~\bibnamefont {{Berger}}}, \bibinfo {author} {\bibfnamefont
  {A.}~\bibnamefont {{Magrez}}}, \bibinfo {author} {\bibfnamefont
  {C.}~\bibnamefont {{Bernhard}}}, \bibinfo {author} {\bibfnamefont
  {M.}~\bibnamefont {{Grioni}}}, \bibinfo {author} {\bibfnamefont
  {A.}~\bibnamefont {{Rubio}}}, \ and\ \bibinfo {author} {\bibfnamefont
  {M.}~\bibnamefont {{Chergui}}},\ }\href@noop {} {\bibfield  {journal}
  {\bibinfo  {journal} {ArXiv e-prints}\ } (\bibinfo {year} {2016})},\ \Eprint
  {http://arxiv.org/abs/1601.01244} {arXiv:1601.01244 [cond-mat.mtrl-sci]}
  \BibitemShut {NoStop}%
\bibitem [{\citenamefont {{Baldini}}\ \emph {et~al.}(2017)\citenamefont
  {{Baldini}}, \citenamefont {{Mann}}, \citenamefont {{Benfatto}},
  \citenamefont {{Cappelluti}}, \citenamefont {{Acocella}}, \citenamefont
  {{Silkin}}, \citenamefont {{Eremeev}}, \citenamefont {{Kuzmenko}},
  \citenamefont {{Borroni}}, \citenamefont {{Tan}}, \citenamefont {{Xi}},
  \citenamefont {{Zerbetto}}, \citenamefont {{Merlin}},\ and\ \citenamefont
  {{Carbone}}}]{Baldini:2017aa}%
  \BibitemOpen
  \bibfield  {author} {\bibinfo {author} {\bibfnamefont {E.}~\bibnamefont
  {{Baldini}}}, \bibinfo {author} {\bibfnamefont {A.}~\bibnamefont {{Mann}}},
  \bibinfo {author} {\bibfnamefont {L.}~\bibnamefont {{Benfatto}}}, \bibinfo
  {author} {\bibfnamefont {E.}~\bibnamefont {{Cappelluti}}}, \bibinfo {author}
  {\bibfnamefont {A.}~\bibnamefont {{Acocella}}}, \bibinfo {author}
  {\bibfnamefont {V.~M.}\ \bibnamefont {{Silkin}}}, \bibinfo {author}
  {\bibfnamefont {S.~V.}\ \bibnamefont {{Eremeev}}}, \bibinfo {author}
  {\bibfnamefont {A.~B.}\ \bibnamefont {{Kuzmenko}}}, \bibinfo {author}
  {\bibfnamefont {S.}~\bibnamefont {{Borroni}}}, \bibinfo {author}
  {\bibfnamefont {T.}~\bibnamefont {{Tan}}}, \bibinfo {author} {\bibfnamefont
  {X.}~\bibnamefont {{Xi}}}, \bibinfo {author} {\bibfnamefont {F.}~\bibnamefont
  {{Zerbetto}}}, \bibinfo {author} {\bibfnamefont {R.}~\bibnamefont
  {{Merlin}}}, \ and\ \bibinfo {author} {\bibfnamefont {F.}~\bibnamefont
  {{Carbone}}},\ }\href@noop {} {\bibfield  {journal} {\bibinfo  {journal}
  {ArXiv e-prints}\ } (\bibinfo {year} {2017})},\ \Eprint
  {http://arxiv.org/abs/1701.04795} {arXiv:1701.04795 [cond-mat.supr-con]}
  \BibitemShut {NoStop}%
\bibitem [{\citenamefont {Sensarma}\ \emph {et~al.}(2010)\citenamefont
  {Sensarma}, \citenamefont {Pekker}, \citenamefont {Altman}, \citenamefont
  {Demler}, \citenamefont {Strohmaier}, \citenamefont {Greif}, \citenamefont
  {J\"ordens}, \citenamefont {Tarruell}, \citenamefont {Moritz},\ and\
  \citenamefont {Esslinger}}]{Sensarma:2010aa}%
  \BibitemOpen
  \bibfield  {author} {\bibinfo {author} {\bibfnamefont {R.}~\bibnamefont
  {Sensarma}}, \bibinfo {author} {\bibfnamefont {D.}~\bibnamefont {Pekker}},
  \bibinfo {author} {\bibfnamefont {E.}~\bibnamefont {Altman}}, \bibinfo
  {author} {\bibfnamefont {E.}~\bibnamefont {Demler}}, \bibinfo {author}
  {\bibfnamefont {N.}~\bibnamefont {Strohmaier}}, \bibinfo {author}
  {\bibfnamefont {D.}~\bibnamefont {Greif}}, \bibinfo {author} {\bibfnamefont
  {R.}~\bibnamefont {J\"ordens}}, \bibinfo {author} {\bibfnamefont
  {L.}~\bibnamefont {Tarruell}}, \bibinfo {author} {\bibfnamefont
  {H.}~\bibnamefont {Moritz}}, \ and\ \bibinfo {author} {\bibfnamefont
  {T.}~\bibnamefont {Esslinger}},\ }\href {\doibase 10.1103/PhysRevB.82.224302}
  {\bibfield  {journal} {\bibinfo  {journal} {Phys. Rev. B}\ }\textbf {\bibinfo
  {volume} {82}},\ \bibinfo {pages} {224302} (\bibinfo {year}
  {2010})}\BibitemShut {NoStop}%
\bibitem [{\citenamefont {Lenar\ifmmode \check{c}\else
  \v{c}\fi{}i\ifmmode~\check{c}\else \v{c}\fi{}}\ and\ \citenamefont
  {Prelov\ifmmode~\check{s}\else
  \v{s}\fi{}ek}(2014)}]{Lenarifmmode-celse-cfiiifmmode-celse-cfi:2014aa}%
  \BibitemOpen
  \bibfield  {author} {\bibinfo {author} {\bibfnamefont {Z.}~\bibnamefont
  {Lenar\ifmmode \check{c}\else \v{c}\fi{}i\ifmmode~\check{c}\else
  \v{c}\fi{}}}\ and\ \bibinfo {author} {\bibfnamefont {P.}~\bibnamefont
  {Prelov\ifmmode~\check{s}\else \v{s}\fi{}ek}},\ }\href {\doibase
  10.1103/PhysRevB.90.235136} {\bibfield  {journal} {\bibinfo  {journal} {Phys.
  Rev. B}\ }\textbf {\bibinfo {volume} {90}},\ \bibinfo {pages} {235136}
  (\bibinfo {year} {2014})}\BibitemShut {NoStop}%
\bibitem [{\citenamefont {Freericks}\ \emph {et~al.}(2009)\citenamefont
  {Freericks}, \citenamefont {Krishnamurthy},\ and\ \citenamefont
  {Pruschke}}]{Freericks:2009aa}%
  \BibitemOpen
  \bibfield  {author} {\bibinfo {author} {\bibfnamefont {J.~K.}\ \bibnamefont
  {Freericks}}, \bibinfo {author} {\bibfnamefont {H.~R.}\ \bibnamefont
  {Krishnamurthy}}, \ and\ \bibinfo {author} {\bibfnamefont {T.}~\bibnamefont
  {Pruschke}},\ }\href {\doibase 10.1103/PhysRevLett.102.136401} {\bibfield
  {journal} {\bibinfo  {journal} {Phys. Rev. Lett.}\ }\textbf {\bibinfo
  {volume} {102}},\ \bibinfo {pages} {136401} (\bibinfo {year}
  {2009})}\BibitemShut {NoStop}%
\bibitem [{\citenamefont {Werner}\ and\ \citenamefont
  {Eckstein}(2013)}]{Werner:2013uq}%
  \BibitemOpen
  \bibfield  {author} {\bibinfo {author} {\bibfnamefont {P.}~\bibnamefont
  {Werner}}\ and\ \bibinfo {author} {\bibfnamefont {M.}~\bibnamefont
  {Eckstein}},\ }\href {\doibase 10.1103/PhysRevB.88.165108} {\bibfield
  {journal} {\bibinfo  {journal} {Phys. Rev. B}\ }\textbf {\bibinfo {volume}
  {88}},\ \bibinfo {pages} {165108} (\bibinfo {year} {2013})}\BibitemShut
  {NoStop}%
\bibitem [{\citenamefont {Strand}(2014)}]{Strand:2014ab}%
  \BibitemOpen
  \bibfield  {author} {\bibinfo {author} {\bibfnamefont {H.~U.~R.}\
  \bibnamefont {Strand}},\ }\href {\doibase 10.1103/PhysRevB.90.155108}
  {\bibfield  {journal} {\bibinfo  {journal} {Phys. Rev. B}\ }\textbf {\bibinfo
  {volume} {90}},\ \bibinfo {pages} {155108} (\bibinfo {year}
  {2014})}\BibitemShut {NoStop}%
\end{thebibliography}%

\appendix

\section{Local many-body view on relaxation}
\label{app:PPRelax}

%
%
\begin{figure}
  \includegraphics[scale=1]
  {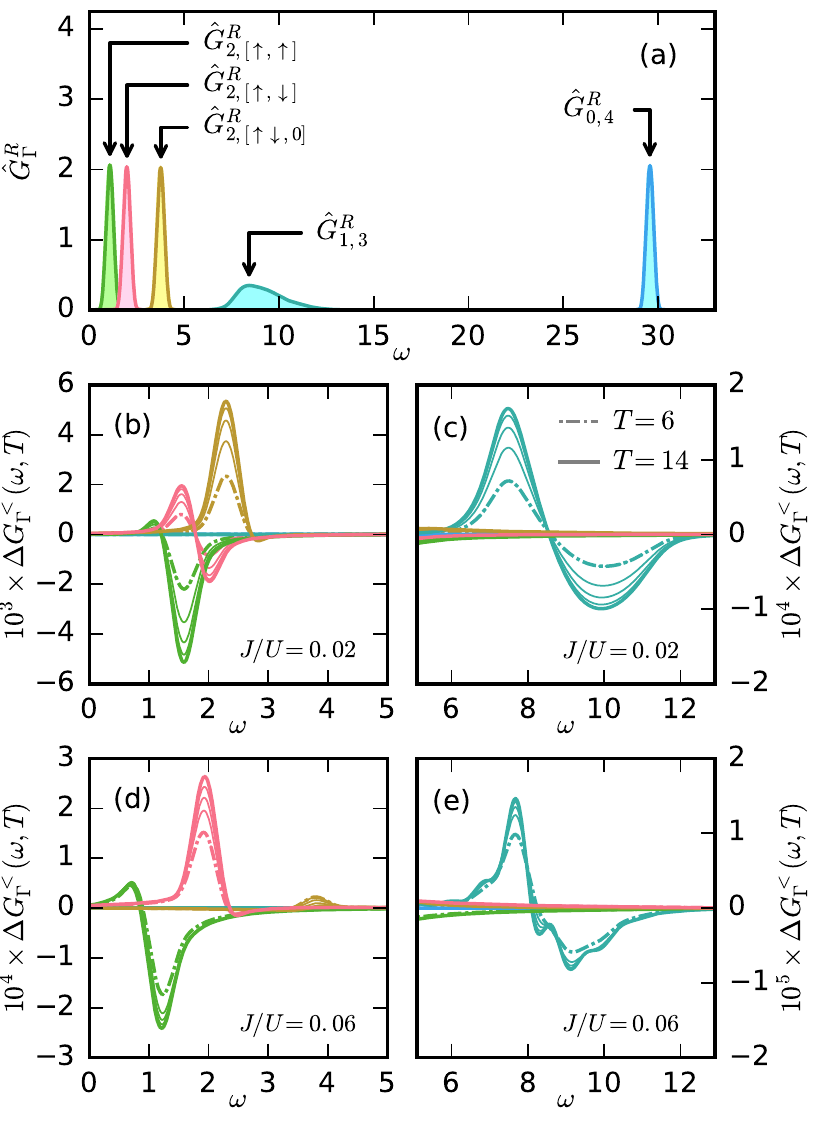} \\[-5mm]
  \caption{\label{fig:ppA}(Color online) Equilibrium pseudo-particle frequency Green's function $\hat{G}^R_\Gamma(\omega)$ for $U=15$, $\beta=1$, $J/U=0.06$ (panel a). Panels (b) to (e): Change in the lesser pseudo-particle Green's function, $\Delta G^<_\Gamma(\omega, T) = G^<_\Gamma(\omega, T) - G^<_\Gamma(\omega, 4)$, relative to absolute time $T=4$ for times $T=6,8,10,12,14$, for $J/U=0.02$ (panels b and c) and $J/U=0.06$ (panels d and e).
  }
\end{figure}
%

The pseudo-particle approach does not only give access to the local many-body density matrix, but also to spectroscopic information on the local many-body states (the pseudo-particles). The equilibrium pseudo-particle spectral functions $\hat{G}_{\Gamma}(\omega, t)$ for all local many-body states $\Gamma$ are shown in Fig.~\ref{fig:ppA}a. The three doublon states, the high-spin $S=1$ state (green), and the high (yellow) and low energy (red) $S=0$ doublon states (split by $J$ and $3J$) are the lowest energy pseudo-particles at half-filling. The singlon-triplon states are intermediate in energy (at $\omega \approx 10$) and the holon-quadruplon states (at $\omega \approx 30$) are the highest energy local many-body states. The singlon-triplon states are the only ones that are dispersive, with a smeared-out spectral distribution. All other states, the doublons and holon-quadruplons are delta-peak like resonances (the peak-widths in Fig.~\ref{fig:ppA} are due to a limited frequency resolution).

Since only the singlon-triplon states are dispersive, kinetic energy relaxation is restricted to these states. Hence, the singlon-triplon scattering processes (IIa \& IIb), schematically shown in Fig.~\ref{fig:Schematic}, are the relevant processes for relaxing excess kinetic energy in the photo-doped Mott insulator. This can also be seen in the time-dependent change in the singlon-triplon occupied density of states [Fig.~\ref{fig:ppA}c], where spectral density is redistributed with time from high to low frequencies within the singlon-triplon pseudo-particle band. As previously noted the speed of the kinetic relaxation is controlled by the Hund's coupling $J$ and comparing $J/U= 0.02$ and $0.06$ the relaxation is only visible in the time range $t = 6$ to $14$ for the lower value of $J$, compare Figs.~\ref{fig:ppA}c and \ref{fig:ppA}e (note the order-of-magnitude scale difference between panels b \& d, and panels c \& e).

The redistribution within the doublon sector can also be analyzed in terms of the time-dependent pseudo-particle spectral function, see Figs.~\ref{fig:ppA}b and \ref{fig:ppA}d.
For $J/U=0.02$ [Fig.~\ref{fig:ppA}b] we observe the redistribution from the high spin $S=1$ doublon (green) to the high energy $S=0$ doublon (yellow), while the spectral weight of the low energy $S=0$ doublon (red) shifts down in energy. For $J/U=0.06$ [Fig.~\ref{fig:ppA}d] the situation is markedly different. The high energy $S=0$ doublon (yellow) is now well separated from the $S=1$ and low energy $S=0$ doublons (green and red) and only a conversion between the last two occurs on the timescale of the calculations. This is a prime example of \emph{kinetic freezing} of the high energy doublon state.

We also note that the pseudo-particle spectral function enables a qualitative understanding of the single-particle spectral function. In particular, the three peak structure in the upper Hubbard band of Fig.~\ref{fig:spA}g, corresponding the removal processes $|3\rangle \rightarrow |2,\Gamma \rangle$, can be understood as the convolution of the lesser triplon spectral function $\hat{G}_{0,3}$ and the retarded doublon spectral functions $\hat{G}_{2,\Gamma}$.

\end{document}